\documentclass[notitlepage,twocolumn,letterpaper,natbib,aps,prd,amsmath,amsfonts,nofootinbib,preprintnumbers,superscriptaddress,secnumarabic]{revtex4-1}
\pdfoutput=1

\usepackage{amssymb,amsmath,latexsym,mathrsfs}
\usepackage{physics}
\usepackage{mathtools}
\usepackage{graphicx}

\usepackage{dcolumn}
\usepackage{multirow}
\usepackage{booktabs}
\usepackage{algorithm}
\usepackage{algpseudocode}
\usepackage{adjustbox}
\usepackage[shortlabels]{enumitem}
\usepackage{bm}
\usepackage[usenames,dvipsnames]{color}
\usepackage[breaklinks,colorlinks,urlcolor=magenta,citecolor=magenta,linkcolor=magenta]{hyperref}
\usepackage{xcolor}
\usepackage{tikz}
\usetikzlibrary{positioning}
\usetikzlibrary{shapes.geometric}
\usetikzlibrary{decorations.pathreplacing,calligraphy}

\definecolor{citecolor}{RGB}{66, 117, 176} 
\definecolor{linkcolor}{RGB}{174, 60, 60} 
\hypersetup{
  linkcolor  = linkcolor,
  citecolor  = citecolor,
  urlcolor   = linkcolor,
  colorlinks = true
}

\begin{document}

\preprint{}

\title{Scalable inference with Autoregressive Neural Ratio Estimation}

\author{Noemi Anau Montel}
\email[ ]{n.anaumontel@uva.nl}
\affiliation{GRAPPA (Gravitation Astroparticle Physics Amsterdam), University of Amsterdam, Science Park 904, 1098 XH Amsterdam, The Netherlands}

\author{James Alvey}
\email[ ]{j.b.g.alvey@uva.nl}
\affiliation{GRAPPA (Gravitation Astroparticle Physics Amsterdam), University of Amsterdam, Science Park 904, 1098 XH Amsterdam, The Netherlands}

\author{Christoph Weniger}
\email[ ]{c.weniger@uva.nl}
\affiliation{GRAPPA (Gravitation Astroparticle Physics Amsterdam), University of Amsterdam, Science Park 904, 1098 XH Amsterdam, The Netherlands}

\begin{abstract}
    \noindent In recent years, there has been a remarkable development of simulation-based inference (SBI) algorithms, and they have now been applied across a wide range of astrophysical and cosmological analyses. There are a number of key advantages to these methods, centred around the ability to perform scalable statistical inference without an explicit likelihood. In this work, we propose two technical building blocks to a specific sequential SBI algorithm, truncated marginal neural ratio estimation (TMNRE). In particular, first we develop autoregressive ratio estimation with the aim to robustly estimate correlated high-dimensional posteriors. Secondly, we propose a slice-based nested sampling algorithm to efficiently draw both posterior samples and constrained prior samples from ratio estimators, the latter being instrumental for sequential inference. To validate our implementation, we carry out inference tasks on three concrete examples: a toy model of a multi-dimensional Gaussian, the analysis of a stellar stream mock observation, and finally, a proof-of-concept application to substructure searches in strong gravitational lensing. In addition, we publicly release the code for both the autoregressive ratio estimator and the slice sampler.
\end{abstract}

\maketitle


\section{Introduction} \label{sec:intro}

\noindent Our understanding of the universe has the potential to be revolutionised by the exponentially growing influx of high quality data from current and upcoming astrophysical or cosmological surveys~\citep{snowmass2021}. Data from current and near-future facilities (\textit{e.g.}~Euclid~\citep{laureijs2011euclid}, JWST \citep{JWST2006}, Rubin-LSST \citep{lsst}, ELT \citep{elt}, Gaia \citep{gaia}, SKA \citep{SKA}, CTA \citep{CTA}) will exceed the peta- and exabyte threshold. Given this status, it is crucial for the community to develop innovative data analysis pipelines, statistical algorithms, data compression techniques, and search pipelines. These will need to handle not only the escalating amount of data, but also its increasing resolution, which directly translates into growing model complexity.

In general, to obtain information about physical models from the data, one has to solve the ``inverse problem", \textit{i.e.}~given an observation, the goal is to infer the parameters of a specific model (or set of models) that are most likely to have generated it. The main tools to solve inverse problems for modern astrophysical and cosmological data analysis have been sampling-based inference methods like Markov-chain Monte Carlo (MCMC) \citep{Metropolis2004, Hastings1970} and nested sampling \citep{Skilling_2006_ns, MultiNest, Handley_2015_polychord} techniques. However, these methods often rely on approximate likelihoods, and the time needed to reach convergence scales poorly with the dimensionality of the explored parameter space. More modern methods are taking up this latter challenge, including gradient-based algorithms such as Hamiltonian Monte-Carlo~\citep{Duane:1987aaa}, or slice-sampling techniques~\citep{Neal_2000,Handley_2015_polychord}.

Novel techniques in the field of simulation-based inference (SBI) are starting to overcome these challenges (for a recent review see \cite{Cranmer_2020}), and have recently gained significant popularity due to a number of appealing features (see \textit{e.g.}~\cite{Alsing_2018, Alsing_2019_cosmo, Lemos_2022, Dax_2021, Wagner_Carena_2021, coogan2022walks, Legin_2022, alvey2023albatross, bhardwaj2023peregrine, Karchev_2022, brehmer2020simulationbased, Zhao_2022, Cole_2021_planck} for examples of method development and applications across cosmology and astroparticle physics). First of all, SBI methods do not require an explicit model of the data likelihood, but instead access its information implicitly via a stochastic simulator, which maps input parameters to data. Secondly, SBI techniques are able to directly target marginal posteriors for the parameters of interest \citep{Alsing_2019, jeffrey2020solving}, which typically improves scalability with the dimensionality of the parameter space. This is because an arbitrarily large number of nuisance parameters can be included whilst targeting the same set of parameters of interest.\footnote{It is important to note that if all parameters contribute equally to the data variance, the implicit data distribution will become noise-dominated. Thus, when referring to scaling to arbitrary number of variables, the data variance is implicitly kept fixed. This limit remains a challenge for sampling-based methods, but is tractable in the SBI framework.} Lastly, sequential SBI approaches that continuously use the acquired inference knowledge to guide the simulator into the relevant part of the parameter space through active learning, have been shown to be particularly simulation efficient \citep{Lueckmann_2021}.

Whilst there are a wide range of SBI algorithms (we refer the reader to Sec.~\ref{subsec:sbi} for a broad overview), the main focus of this work will be truncated marginal neural ratio estimation (TMNRE) \citep{Miller_2020, Miller_2021}. This is a sequential implementation of the general neural ratio estimation technique \citep{hermans2020likelihoodfree} that composes well with marginalisation. In essence, neural ratio estimation trains a neural network to approximate the posterior-to-prior ratio by solving a binary classification problem (for more details see Sec.~\ref{subsec:nre}).
TMNRE has been shown to be successful in a number of different physics applications, such as cosmic microwave background analyses \citep{Cole_2021_planck}, strong lensing image analysis \citep{Anau_Montel_2022, coogan2022walks}, supernovae Ia cosmology \citep{Karchev_2022}, gravitational wave parameter inference \citep{bhardwaj2023peregrine, alvey2023things}, along with other applications \citep{Hartman_2023_swyft, saxena2023constraining, montel2022detection, alvey2023albatross}.

So far, TMNRE applications have focused on estimating low-dimensional ($d \lesssim 2$) marginal posteriors for the parameters of interest. This choice was motivated by the fact that directly targeting these low-dimensional marginal posteriors alleviates sampling problems associated with high dimensionality, whilst still maintaining much of the information relevant for scientific conclusions. 
Nevertheless, there are situations when focusing on low-dimensional marginal posterior estimates makes TMNRE inherently inefficient, for example when the marginals of interest are highly correlated (or multi-modal). In these cases it is therefore necessary to produce accurate high-dimensional joint estimates to account for these correlations, since higher-dimensional structure can be obscured by low-dimensional projections. This limitation becomes especially relevant for the sequential aspect of the algorithm, which cannot leverage the information regarding the correlations while guiding the simulations. However, the necessity for high-dimensional joint estimates clashes with one know failure mode of density ratio estimators, the so called ``density-chasm problem", described in \cite{rhodes2020telescoping}. In essence, density ratio estimators can fail whenever the gap (in the sense of \citet{rhodes2020telescoping}) between the two densities is large, since the binary classifier can obtain almost perfect accuracy with a relatively poor estimate of the density ratio. This problem is exacerbated in high-dimensions. 

The present paper is a step forward towards addressing these competing limitations. In this work, we propose two new building blocks of the TMNRE framework: an autoregressive implementation of neural ratio estimation for scalable, high-dimensional (marginal) posterior inference, and a slice-based nested sampling algorithm to efficiently draw not only posterior samples, but also constrained prior samples. The latter set of samples, as will be discussed in Sec.~\ref{subsec:sequential}, are necessary for TMNRE's implementation of active learning. 

In recent years, autoregressive models have shown great potential in scaling to high-dimensional distribution estimation problems, see \textit{e.g.}~\cite{Germain_2015, Uria_2016, Papamakarios_2018_masked}. The term ``autoregressive" originates from the time-series modeling literature, where it refers to the practice of using previous time-step observations to predict the value at the current time step. Autoregressive models function in a similar fashion by decomposing a $d$-dimensional joint density into a product of $d$ 1-dimensional conditional distributions, as in Eq.~\eqref{eq:autoregressive}. An autoregressive model is then defined by the parameterisation of all $d$ conditionals.

We introduce the slice sampling component because neural ratio estimators do not come with sampling functionalities. In general, to sample from the estimated (marginal) posterior, we require Monte Carlo sampling algorithms, especially in high dimensions. Moreover, we are interested in obtaining constrained prior samples for the purposes of active learning. In this context, whilst nested sampling was primarily born as a general purpose integration algorithm in high-dimensions, it has primarily been applied to perform Bayesian inference, in particular to estimate the Bayesian evidence and the posterior. Interestingly, at the heart of nested sampling algorithms resides also the problem of constrained prior sampling \citep{Ashton_2022}. Motivated by the success of slice-based nested samplers \citep{Neal_2000, Handley_2015_polychord}, we implement a custom slice sampler to draw \emph{both} posterior and constrained prior samples.  

\vspace{10pt}
\noindent The structure of this work is as follows. We begin in Sec.~\ref{sec:background} by laying out relevant technicalities in the pedagogical background. We then introduce this work's main contribution: autoregressive neural ratio estimation and prior truncation through slice-based nested sampling, in Sec.~\ref{sec:method}. In Sec.~\ref{sec:experiments}, we present results on a toy example, and applications to stellar streams and substructure parameter inference in strong gravitational lensing images, which initially motivated the development of the presented tools. Finally, we provide some discussion regarding possible limitations and outlook, concluding in Sec.~\ref{sec:conclusion}. 

\vspace{10pt}
\noindent \textbf{Code.} We release a publicly available implementation of the autoregressive ratio estimator model in the TMNRE package \texttt{swyft}\footnote{\url{https://github.com/undark-lab/swyft}}, and an implementation of the slice-based nested sampler in \texttt{torchns}\footnote{\url{https://github.com/undark-lab/torchns}}.

\section{Technical background} \label{sec:background}

\noindent The goal of this section is to provide some brief technical background for the main contributions and results given in Secs.~\ref{sec:method} and~\ref{sec:experiments}. More specifically, we will first give a brief overview of SBI and its various implementations. We will then focus our attention on the particular SBI technique we use in this work, \emph{neural ratio estimation}, and present the key features of the algorithm. Finally, we will discuss different implementations of active learning that can be used for sequential inference.

\subsection{Simulation-based inference} \label{subsec:sbi}
\noindent In scientific analyses, inferring the probability distribution of model parameters $\boldsymbol{\theta}$ for a given observation $\mathbf x_o$ is a ubiquitous task. In a Bayesian framework, the posterior distribution for model parameters $\boldsymbol{\theta}$ follows from Bayes' theorem
\begin{equation} \label{eq:bayes}
    p(\boldsymbol{\theta}\mid\mathbf{x})=\cfrac{p(\mathbf{x}\mid\boldsymbol{\theta})}{p(\mathbf{x})} \, p(\boldsymbol{\theta}) \, ,
\end{equation}
where $p(\mathbf{x}\mid\boldsymbol{\theta})$ is the likelihood of the data $\mathbf{x}$ for given parameters $\boldsymbol{\theta}$, $p(\boldsymbol{\theta})$ is the prior probability distribution over the parameters, and $p(\mathbf{x})$ is the evidence of the data. As mentioned in Sec.~\ref{sec:intro}, SBI algorithms do not explicitly calculate the likelihood function, but instead rely on a stochastic simulator that maps from model parameters $\boldsymbol{\theta}$ to data $\mathbf{x}$. This mapping is equivalent to sampling from the distribution $\mathbf{x} \sim p(\mathbf{x} \mid \boldsymbol{\theta})$, which is effectively an implicit representation of the likelihood. As a result, one is able to shift the focus towards building a realistic simulator/forward model. In principle, this approach allows for the simultaneous inclusion of all relevant processes that can affect the data, regardless of whether a full probabilistic description is tractable or not.

Turning more towards the history and development of simulation-based inference, the first established SBI technique was approximate Bayesian computation (ABC). ABC is a rejection sampling algorithm where proposed samples $\mathbf x$ from the forward model are compared to the target observed data $\mathbf x_o$ and accepted based on a user defined metric and acceptance tolerance \citep{sisson2018overview}. One of the explicit disadvantages of ABC, however, is that it requires hand-crafted distance measures over summary statistics to compare simulations to the data. It also typically requires a relatively large -- \textit{e.g.} as compared to other traditional methods -- simulation budget to reach convergence.

Moving beyond this classical SBI approach, in recent years, a number of new neural network-based SBI techniques have been proposed. This is largely thanks to the advances in deep learning and automatic differentiation which have allowed for the processing of significantly more complex data structures (such as images), as well as the efficient optimisation of parametric functions (here, neural networks) with respect to some custom loss. 

Bayes' theorem in Eq.~\eqref{eq:bayes} hints at a few different approaches for the implementation of neural SBI algorithms, as reviewed in \cite{Cranmer_2020, Lueckmann_2021}. Broadly, the taxonomy of neural SBI algorithms include neural posterior estimation (NPE) which employs density estimation techniques to directly estimate the posterior \citep{Papamakarios_2018_density, greenberg2019automatic}; neural likelihood estimation (NLE) which instead uses density estimation to learn an approximation to the likelihood \citep{papamakarios2019sequential}; and neural ratio estimation (NRE) \citep{hermans2020likelihoodfree,Miller_2020} which uses classifiers to approximate the likelihood-to-evidence or posterior-to-prior ratio. In this work we will focus on NRE and its extensions.

\subsection{Neural Ratio Estimation} \label{subsec:nre}
\noindent Following \cite{hermans2020likelihoodfree}, neural ratio estimation rephrases posterior estimation as a binary classification problem. Given a simulator $p(\mathbf{x}, \boldsymbol{\theta}) = p(\mathbf{x}\mid\boldsymbol{\theta})p(\boldsymbol{\theta})$, the idea behind NRE is to train a binary classifier to distinguish between data and parameter pairs drawn jointly $\mathbf{x}, \boldsymbol{\theta} \sim p(\mathbf{x}, \boldsymbol{\theta})$ or marginally $\mathbf{x}, \boldsymbol{\theta} \sim p(\mathbf{x}) p(\boldsymbol{\theta})$.\footnote{Throughout this work we will refer to joint samples as \emph{positive} training examples, and to marginal samples as \emph{negative} training examples.} More precisely, the binary classifier $d_\mathbf{\phi}(\mathbf{x}, \boldsymbol{\theta})$ estimates the probability that a data-parameter pair $(\mathbf{x}, \boldsymbol{\theta})$ is drawn jointly
\begin{equation} \label{eq:classifier}
    d_\mathbf{\phi}(\mathbf{x}, \boldsymbol{\theta}) = \frac{p(\mathbf{x}, \boldsymbol{\theta})}{p(\mathbf{x}, \boldsymbol{\theta}) + p(\mathbf{x}) p(\boldsymbol{\theta})} \equiv \sigma[ \log r_\mathbf{\phi}(\boldsymbol{\theta};\mathbf{x})] \, ,
\end{equation}
where we have introduced the neural network parameters $\mathbf{\phi}$ and the sigmoid function $\sigma(y) \equiv 1 / (1 + e^{-y})$. We have also introduced the posterior-to-prior ratio, likelihood-to-evidence ratio, or joint-to-marginal distribution ratio:
\begin{equation} \label{eq:ratio}
    r(\boldsymbol{\theta};\mathbf{x}) \equiv \frac{p(\boldsymbol{\theta} \mid \mathbf{x})}{p(\boldsymbol{\theta})} = \frac{p(\mathbf{x} \mid \boldsymbol{\theta})}{p(\mathbf{x})} =  \frac{p(\mathbf{x}, \boldsymbol{\theta})}{p(\mathbf{x}) p(\boldsymbol{\theta})} \, ,
\end{equation}
which are all equivalent as a result of conditional probability rules.
If we can train a ratio estimator $r_\phi$ via this supervised classification task, then, knowing the prior distribution, we can obtain an estimate of the posterior by weighting prior samples as $p(\boldsymbol{\theta}\mid\mathbf{x}) = r_\phi(\boldsymbol{\theta}; \mathbf{x})p(\boldsymbol{\theta})$.

One way to perform this training is to minimise the binary cross-entropy loss
\begin{equation}\label{eq:loss}
\begin{split}
    \mathcal{L}[r_\mathbf{\phi}(\boldsymbol{\theta};\mathbf{x})] = &-\int \mathrm{d} \mathbf{x} \mathrm{d} \boldsymbol{\theta} \left\{ p(\mathbf{x}, \boldsymbol{\theta}) \log \sigma[ \log r_\mathbf{\phi}(\boldsymbol{\theta};\mathbf{x}) ] \right. \\
    & \left. + p(\mathbf{x}) p(\boldsymbol{\theta}) \log\left[ 1 - \sigma[ \log r_\mathbf{\phi}(\boldsymbol{\theta};\mathbf{x}) ] \right] \right\},
\end{split}
\end{equation}
via stochastic gradient descent, and find the optimal parameters $\phi$ of the ratio estimator. Importantly, one can also analytically minimise this loss and show that the optimal classifier is indeed $d_\mathbf{\phi}(\mathbf{x}, \boldsymbol{\theta}) = \sigma[ \log r(\boldsymbol{\theta};\mathbf{x})]$, meaning that we can directly obtain the posterior-to-prior ratio through this optimisation process. In practice, both the data and the parameters are typically high-dimensional objects, and therefore neural network-based classifiers that are able to process this complex data and be efficiently optimised are a necessary choice.

\vspace{10pt}
\noindent Given this general NRE setup, it is straightforward to extend ratio estimation into \emph{marginal} ratio estimation \citep{Miller_2020} and \emph{conditional} ratio estimation. 

\vspace{10pt}
\noindent \textbf{Marginal Ratio Estimation.} It has been shown that significant simulation efficiency can be achieved by directly targeting marginal posteriors, rather than marginalising over samples from the full joint distribution \citep{Cole_2021_planck, bhardwaj2023peregrine}. To see how we can perform marginal inference with NRE, consider a model with a full joint distribution $p(\mathbf{x},\boldsymbol{\vartheta}, \boldsymbol{\eta})= p(\mathbf{x}\mid\boldsymbol{\vartheta}, \boldsymbol{\eta})p(\boldsymbol{\vartheta}, \boldsymbol{\eta})$, where we have split the set of all model parameters $\boldsymbol{\theta} = \{\boldsymbol{\vartheta},\boldsymbol{\eta}\}$ into parameters of interest $\boldsymbol{\vartheta}$ (which we want to infer) and nuisance parameters $\boldsymbol{\eta}$ (which we want to marginalise over).
For marginal posterior estimation, nuisance parameters must be sampled, but not passed to the binary classifier. More specifically, the marginal ratio for parameters of interest $\boldsymbol{\vartheta}$,
\begin{equation}
    \label{eq:ratio_marginal}
    r(\boldsymbol{\vartheta};\mathbf{x}) \equiv \frac{p(\boldsymbol{\vartheta} \mid \mathbf{x})}{p(\boldsymbol{\vartheta})} =  \int \mathrm{d}\boldsymbol{\eta} \, \frac{p(\boldsymbol{\theta, \eta} \mid \mathbf{x})}{p(\boldsymbol{\vartheta})} p(\boldsymbol{\eta})  \, ,
\end{equation}
will be automatically learned by the binary classifier trained on $(\mathbf{x}, \boldsymbol{\vartheta})$ pairs. This implicitly marginalises over nuisance parameters $\boldsymbol{\eta}$, since simulations $\mathbf x \sim p(\mathbf x | \boldsymbol{\vartheta}, \boldsymbol{\eta})$ will contain their variance. 

\vspace{10pt}
\noindent \textbf{Conditional Ratio Estimation.} Within the NRE framework, it is also possible to perform conditional inference. In this case, the binary classifier must be informed about the value of the parameters we want to condition over. More specifically, in order to learn the conditional posterior for parameter $\vartheta_i$ given parameter $\vartheta_j$ (with $j \neq i$), the binary classifier will be shown as positive training example $\mathbf{x}, \vartheta_j, \vartheta_i \sim p(\mathbf{x}, \vartheta_j, \vartheta_i)$ and as negative training examples $\mathbf{x}, \vartheta_j, \vartheta_i \sim p(\mathbf{x}, \vartheta_j)p(\vartheta_i)$. We can then express the conditional ratio trained on $(\{\mathbf{x}, \vartheta_j\}, \vartheta_i)$ pairs as
\begin{equation}
    \label{eq:ratio_conditional}
    r(\vartheta_i;\mathbf{x}, \vartheta_j) \equiv
    \frac{p(\vartheta_i \mid \mathbf x, \vartheta_j)}{ p(\vartheta_i)} \, .
\end{equation}
It is possible to condition more than one parameter $\vartheta_i$ on more than one variable $\vartheta_j$ just by correctly passing to the classifier positive and negative training examples as defined above. In Sec.~\ref{subsec:anre}, we will use 1-dimensional versions of these ratio estimators, conditioned on multiple variables, to implement autoregressive NRE.

\subsection{Sequential methods} \label{subsec:sequential}

\noindent The taxonomy of SBI algorithms can be refined further based on whether they use active learning to guide the simulator towards parts of the parameter space that are most relevant for a specific target observation $\mathbf x_o$. In particular, \emph{sequential} SBI approaches adaptively choose informative simulations by using sequentially refined proposal distributions for the model parameters. They have been reported to outperform and be more simulation-efficient with respect to non-sequential ones across a number of different benchmark tasks \citep{Lueckmann_2021}. 

The core intuition behind sequential SBI algorithms is the following: given a single observation of interest, $\mathbf x_o$, sampling parameters from the entire prior space to generate training data may not be efficient, since it leads to training data $(\mathbf x, \boldsymbol \vartheta)$ that has significant variance compared to the target observation $\mathbf x_o$. Therefore, for a fixed simulation budget, the training samples contain only limited information about the posterior $p(\boldsymbol \vartheta\mid\mathbf x_o)$. The alternative proposed by sequential algorithms, in order to increase simulation efficiency, is to draw parameters from an adaptive proposal distribution $\tilde p_{R}(\boldsymbol \vartheta)$, resulting in training data that matches the observation of interest more closely in each sequential round $R$. 

The classification of sequential SBI algorithms can then be further sifted based on how they acquire new, informative simulations. In particular, the sequential techniques adopted in \cite{Papamakarios_2018_density, greenberg2019automatic, papamakarios2019sequential, hermans2020likelihoodfree, durkan2020contrastive} draw new simulations for the next round from the \emph{approximate posteriors} learned in each round. However, this approach suffers from two limitations. First, several frequently used diagnostic tools for SBI depend on performing inference across multiple observations (\textit{e.g.}~expected coverage tests \citep{hermans2022trust}). In this setting, to perform these tests, one would have to generate new simulations and network retraining for each observation, which is often prohibitively expensive. Second, marginal posterior estimation is in general affected by the proposal distribution, since one implicitly integrates over it. As a result, this approach is unsuitable for learning multiple marginal posteriors simultaneously over a number of sequential rounds (for a possible workaround see \cite{Alsing_2019_fast}).

To overcome the limitations of this sequential scheme, \cite{Miller_2021} proposed a hard-likelihood \emph{prior truncation} scheme, applicable to NRE, that composes well with marginalisation and is locally amortised.\footnote{With \emph{locally amortised} inference we refer to an inference that can be repeated several times, without retraining, with distinct observations that live in the support of the truncated prior.}
This prior truncation scheme iteratively discards in rounds $R$ low likelihood(-to-evidence) regions, where the current approximate likelihood-to-evidence ratio evaluated for the target observation is below a user defined threshold $\epsilon$. Concretely, this means keeping the region of parameter space defined by
\begin{equation}\label{eq:gamma_r}
    \Gamma^{(R)}_{\boldsymbol \vartheta} = \{ \boldsymbol \vartheta \in \mathbb R^d : r^{(R)}(\boldsymbol \vartheta ; \mathbf x) > \epsilon\} \;,
\end{equation}
where $d$ is the dimensionality of the $\boldsymbol \vartheta$ parameter space. Similar truncated proposals have also been introduced in \cite{deistler2022truncated} in the context of NPE, where the condition is instead on the current estimated posterior, \textit{i.e.}~$p^{(R)}(\boldsymbol \vartheta \mid \mathbf x) > \epsilon$.

Importantly, since this sequential scheme does not modify the shape of the prior proposal distribution, but only restricts its support, the inference is still locally amortised in the constrained proposal distribution region.  This enables empirical tests of the inference result (see \textit{e.g.}~\cite{Cole_2021_planck}). Moreover, it is also possible to use the same training data generated for a round to efficiently train arbitrary marginal posteriors for any set of model parameters.

\vspace{10pt}
\noindent To summarise this section, we have briefly discussed the field of SBI techniques, before describing in detail the NRE algorithm and sequential SBI approaches. In doing so, we have explained the technicalities of our particular implementation of SBI, TMNRE, that builds on three key components: active learning through prior truncation (T), focus on marginal inference (M), and NRE.

\section{Methodology: extending TMNRE} \label{sec:method}

\noindent Up to now, applications of TMNRE have typically focused on estimating low-dimensional ($d \lesssim 2$) marginal posteriors. On the other hand, whilst correlations between two parameters can always be estimated by training an appropriate ratio estimator, doing this for all pairwise combinations of a large number of parameters becomes quickly infeasible. Furthermore, even if it were done, two-dimensional correlations do not provide any information about higher order correlations since they are just projections. In certain scenarios, this might be crucial for calibrating the quality and accuracy of the inference methods and strongly motivates the development of techniques able to robustly estimate higher-dimensional posterior distributions.

As discussed in Sec.~\ref{sec:intro}, however, there are challenges related to both the estimation of high-dimensional ratios and subsequently sampling from them. In this work, we propose two new building blocks of the TMNRE framework to address these issues: \emph{autoregressive NRE} for scalable, multi-dimensional (marginal) posterior estimation, and \emph{slice sampling} to efficiently sample from a multi-dimensional posterior and truncated prior.\footnote{This work has been informed by private communications surrounding a complementary paper in preparation \citep{PolySwyft}.}

\subsection{Autoregressive Neural Ratio Estimation} \label{subsec:anre}

\noindent Across the various SBI techniques, autoregressive models have been developed in the context of density estimation algorithms, NPE~\citep{Uria_2016, Papamakarios_2018_masked} and NLE~\citep{papamakarios2019sequential}, and have been shown to be among the best performing density estimators~\citep{Lueckmann_2021}. They have been successfully applied to the Dark Energy Survey and global 21cm signal experiments~\citep{bevins2022removing}, and for deep generative models for galaxy image simulations~\citep{Lanusse_2021}.

In a nutshell, autoregressive models turn the estimation of a $d$-dimensional joint density into the estimation of $d$ 1-dimensional conditional densities using the chain rule of probability
\begin{equation}\label{eq:autoregressive}
    p(\boldsymbol \vartheta \mid\mathbf x) =  p(\vartheta_1\mid\mathbf{x}) \prod_{i=2}^d p(\vartheta_i\mid\mathbf{x}, \boldsymbol \vartheta_{1:i-1})\;,
\end{equation}
where we have introduced the compact notation $\boldsymbol \vartheta_{1:i-1} \equiv \{\vartheta_1, ..., \vartheta_{i-1}\}$. One can thus define an autoregressive model simply by specifying a parameterisation of all $d$ conditionals.

In the context of NRE, our key quantity of interest is the ratio $r(\boldsymbol \vartheta; \mathbf x)$ as defined in Eq.~\eqref{eq:ratio}. One way to estimate this quantity autoregressively is by considering the following components. The first quantity, $A$, estimates the ratio between the joint posterior distribution and the independent marginal priors,
\begin{equation} \label{eq:A}
\begin{split}
    A &= 
    \frac{p(\boldsymbol \vartheta\mid \mathbf x)}{\prod_{i=1}^d p(\vartheta_i)} \\
    & = \frac{p(\vartheta_1\mid \mathbf x)}{p(\vartheta_1)} \prod_{i=2}^d \frac{p(\vartheta_i\mid \mathbf x, \boldsymbol \vartheta_{1:i-1})}{ p(\vartheta_i)} \\
    & = 
    r(\vartheta_1;\mathbf x)  \prod_{i=2}^d  r(\vartheta_i;\mathbf x, \boldsymbol \vartheta_{1:i-1}) \;.
\end{split}
\end{equation}
The second component, $B$, models the dependencies between the model parameters,
\begin{equation} \label{eq:B}
\begin{split}
    B &= 
    \frac{p(\boldsymbol \vartheta)}{\prod_{i=1}^d p(\vartheta_i)} \\
    & = \prod_{i=2}^d \frac{p(\vartheta_i \mid \boldsymbol \vartheta_{1:i-1})}{p(\vartheta_i)} \\
    & = \prod_{i=2}^d r(\vartheta_i;\boldsymbol \vartheta_{1:i-1}) \;.
\end{split}
\end{equation}
Our key quantity of interest is then obtained as 
\begin{equation} \label{eq:AB}
    r(\boldsymbol \vartheta;\mathbf x) = \frac{p(\boldsymbol \vartheta \mid \mathbf x)}{p(\boldsymbol \vartheta)} = A/B \;.
\end{equation}
It is important to highlight the role of the $B$ component in this algorithm. In the case of sequential inference, even if the initial prior distributions from which parameters are drawn are independent, the sequential proposal distribution will account for non-trivial correlations between parameters in the constrained prior region (via the condition in Eq.~\eqref{eq:gamma_r}). It is therefore crucial to properly account for these correlations between parameters through $B$ since they will implicitly be present in the training data. 

Also note that in both the definitions of $A$ and $B$, we have used the notation introduced in Eq.~\eqref{eq:ratio_conditional} for conditional ratios. An alternative formulation of an autoregressive model for ratio estimation that uses a single network to model both components is presented in App.~\ref{apx:anre}.

The key advantage of autoregressive NRE lies in its ability to handle intricate dependencies among variables by focusing on one parameter at a time. In ``vanilla" NRE, for high-dimensional parameter spaces, the discriminating power of the binary classifier can be quickly saturated leading to a poor estimate of the joint density ratio. Here, the full joint distribution is modelled through 1-dimensional conditional ratios, leading to a more stable and accurate estimation of the full joint ratio. However, the presented autoregressive NRE method does inherit the generic drawbacks of autoregressive models. Perhaps the most relevant is their sensitivity to the order of the conditional probabilities, since in practice it is difficult to know which of the factorially many orders is the most efficient in each case \citep{Papamakarios_2018_masked}. A possible solution was presented in \cite{uria2014deep}, that introduced an efficient procedure to simultaneously train an autoregressive model for all possible orderings of the variables. We will study this effect explicitly in the stellar streams example given in Sec.~\ref{subsec:stream}.

\subsection{Prior truncation strategies}\label{subsec:truncation}

\noindent The original prior truncation scheme proposed in the TMNRE formalism \citep{Miller_2021} is a parameter-wise truncation based on 1-dimensional marginal ratio estimators. As a result, the prior gets restricted to a truncation region that has the shape of a hyper-rectangular box (``box" truncation), 
\begin{equation} \label{eq:box}
    \tilde p_R(\boldsymbol \vartheta) = \frac1Z \mathbb{I}(\vartheta_1 \in \Gamma_{\vartheta_1}^{(R-1)}) \times \cdots \times\mathbb{I}(\vartheta_d \in \Gamma_{\vartheta_d}^{(R-1)}) p(\boldsymbol \vartheta) \,,
\end{equation}
where we have introduced the indicator function $\mathbb{I}$ which is unity on the truncated prior support $\Gamma_{\vartheta_i}^R$ and zero otherwise, and the normalizing constant $Z$ which can be interpreted as the fractional volume of the truncated prior.

One of the drawbacks of using this box truncation scheme is that it neglects parameter correlations. For higher ($d \gtrsim 2$) dimensional marginal posteriors, this results in the new constrained region usually containing significantly more probability mass than is actually required (see \textit{e.g.} \cite{Karchev_2022}).

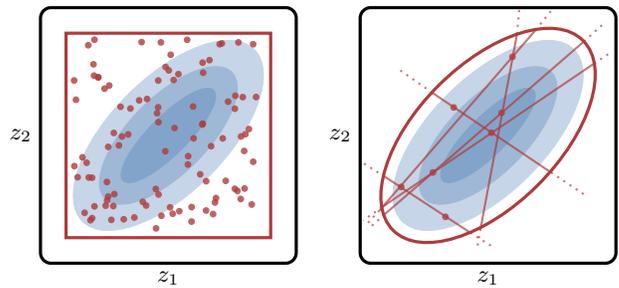
\begin{figure}
\centering
\begin{minipage}{\linewidth}
\begin{tikzpicture}[baseline=(current bounding box.center), scale=0.85]

  \pgfmathsetseed{1}

  \definecolor{myblue}{RGB}{66, 117, 176}
  \definecolor{myred}{RGB}{174, 60, 60}

  \pgfmathsetmacro{\N}{100}

  \draw[opacity=0.3, myblue, fill=myblue, rotate around={45:(-2,-2)}] (-2,-2) ellipse (1.9 and 0.9);
  \draw[opacity=0.3, myblue, fill=myblue, rotate around={45:(-2,-2)}] (-2,-2) ellipse (1.4 and 0.6);
  \draw[opacity=0.3, myblue, fill=myblue, rotate around={45:(-2,-2)}] (-2,-2) ellipse (1 and 0.3);
  
  \draw[very thick, line cap=round, rounded corners=4pt] (-4, -4) rectangle (0, 0);
  \node[below] at (-2, -4) {$z_1$};
  \node[left] at (-4, -2) {$z_2$};

  \draw[very thick, myred] (-3.6, -3.6) rectangle (-0.4, -0.4);

  \foreach \i in {1,...,\N} {
    \pgfmathsetmacro{\x}{-3.5 + rnd * 3.}
    \pgfmathsetmacro{\y}{-3.5 + rnd * 3.}
    \fill[myred, opacity=0.8] (\x, \y) circle (1.5pt);
  }

  \draw[very thick, line cap=round, rounded corners=4pt] (1, -4) rectangle (5, 0);
  \node[below] at (3, -4) {$z_1$};
  \node[left] at (1, -2) {$z_2$};

  \draw[opacity=0.3, myblue, fill=myblue, rotate around={45:(3,-2)}] (3,-2) ellipse (1.9 and 0.9);
  \draw[opacity=0.3, myblue, fill=myblue, rotate around={45:(3,-2)}] (3,-2) ellipse (1.4 and 0.6);
  \draw[opacity=0.3, myblue, fill=myblue, rotate around={45:(3,-2)}] (3,-2) ellipse (1 and 0.3);

  \draw[very thick, myred, rotate around={45:(3,-2)}] (3,-2) ellipse (2.1 and 1.1);

    \pgfmathsetmacro\x{2.93} 
    \pgfmathsetmacro\y{-1.31}
    \fill[myred, opacity=0.8, rotate around={45:(3,-2)}] (\x,\y) circle (1.5pt);

    \pgfmathsetmacro\newx{3.067} 
    \pgfmathsetmacro\newy{-2.007}
    \pgfmathsetmacro\m{-0.65}
    \fill[myred, opacity=0.8, rotate around={45:(3,-2)}] (\newx,\newy) circle (1.5pt);
    \draw[myred, thick, opacity=0.7, rotate around={45:(3,-2)}] (\x-0.075,\y+0.4) -- (\newx+0.22,\newy-1.1);
    \draw[myred, thick, opacity=0.7, dotted, rotate around={45:(3,-2)}] (\x-0.18,\y+1) -- (\x-0.075,\y+0.4);
    \draw[myred, thick, opacity=0.7, dotted, rotate around={45:(3,-2)}] (\newx+0.22,\newy-1.1) -- (\newx+0.34,\newy-1.7);
    \xdef\x{\newx}
    \xdef\y{\newy}

    \pgfmathsetmacro\newx{1.98} 
    \pgfmathsetmacro\newy{-1.80}
    \fill[myred, opacity=0.8, rotate around={45:(3,-2)}] (\newx,\newy) circle (1.5pt);
    \draw[myred, thick, opacity=0.7, rotate around={45:(3,-2)}] (\x+1.9,\y-0.37) -- (\newx-0.95,\newy+0.18);
    \draw[myred, thick, opacity=0.7, dotted, rotate around={45:(3,-2)}] (\x+1.9,\y-0.37) -- (\x+2.2,\y-0.43);
    \draw[myred, thick, opacity=0.7, dotted, rotate around={45:(3,-2)}] (\newx-0.95,\newy+0.18) -- (\newx-1.3,\newy+0.24);
    \xdef\x{\newx}
    \xdef\y{\newy}

    \pgfmathsetmacro\newx{3.4} 
    \pgfmathsetmacro\newy{-1.9}
    \fill[myred, opacity=0.8, rotate around={45:(3,-2)}] (\newx,\newy) circle (1.5pt);
    \draw[myred, thick, opacity=0.7, rotate around={45:(3,-2)}] (\x-1.,\y+0.07) -- (\newx+1.7,\newy-0.12);
    \draw[myred, thick, opacity=0.7, dotted, rotate around={45:(3,-2)}] (\x-1.,\y+0.07) -- (\x-1.35,\y+0.09);
    \draw[myred, thick, opacity=0.7, dotted, rotate around={45:(3,-2)}] (\newx+1.7,\newy-0.12) -- (\newx+2.1,\newy-0.14);
    \xdef\x{\newx}
    \xdef\y{\newy}

    \pgfmathsetmacro\newx{4.14} 
    \pgfmathsetmacro\newy{-1.4}
    \fill[myred, opacity=0.8, rotate around={45:(3,-2)}] (\newx,\newy) circle (1.5pt);
    \draw[myred, thick, opacity=0.7, rotate around={45:(3,-2)}] (\x-1.517,\y-1.05) -- (\newx+0.3,\newy+0.2);
    \draw[myred, thick, opacity=0.7, dotted, rotate around={45:(3,-2)}] (\x-1.517,\y-1.05) -- (\x-1.8,\y-1.25);
    \draw[myred, thick, opacity=0.7, dotted, rotate around={45:(3,-2)}] (\newx+0.3,\newy+0.2) -- (\newx+0.6,\newy+0.45);
    \xdef\x{\newx}
    \xdef\y{\newy}
    
    \pgfmathsetmacro\newx{1.47} 
    \pgfmathsetmacro\newy{-1.61}
    \fill[myred, opacity=0.8, rotate around={45:(3,-2)}] (\newx,\newy) circle (1.5pt);
    \draw[myred, thick, opacity=0.7, rotate around={45:(3,-2)}] (\x+0.55,\y+0.04) -- (\newx-0.5,\newy-0.02);
    \draw[myred, thick, opacity=0.7, dotted, rotate around={45:(3,-2)}] (\x+0.55,\y+0.04) -- (\x+1,\y+0.09);
    \draw[myred, thick, opacity=0.7, dotted, rotate around={45:(3,-2)}] (\newx-0.5,\newy-0.02) -- (\newx-0.8,\newy-0.03);
    \xdef\x{\newx}
    \xdef\y{\newy}

    \pgfmathsetmacro\newx{1.628}
    \pgfmathsetmacro\newy{-2.43}
    \fill[myred, opacity=0.8, rotate around={45:(3,-2)}] (\newx,\newy) circle (1.5pt);
    \draw[myred, thick, opacity=0.7, rotate around={45:(3,-2)}] (\x-0.07,\y+0.3) -- (\newx+0.1,\newy-0.45);
    \draw[myred, thick, opacity=0.7, dotted, rotate around={45:(3,-2)}] (\x-0.07,\y+0.3) -- (\x-0.16,\y+0.7);
    \draw[myred, thick, opacity=0.7, dotted, rotate around={45:(3,-2)}] (\newx+0.1,\newy-0.45) -- (\newx+0.2,\newy-0.9);
    \xdef\x{\newx}
    \xdef\y{\newy}

\end{tikzpicture}
\end{minipage}
\caption{Prior truncation strategies. We visualize in a 2-dimensional parameter space a constrained prior region using hyper-rectangular boxes as defined in Eq.~\eqref{eq:box} (red rectangle on the left), and a constrained prior region using hard iso-likelihood-to-evidence ratio contours as defined in Eq.~\eqref{eq:block} (red ellipse on the right). Sampling from hyper-rectangular boxes is equivalent to uniform sampling, whereas sampling from iso-likelihood-to-evidence ratio contours requires more sophisticated sampling algorithms. In particular, we use a slice-based nested sampler as explained in Sec.~\ref{subsec:ns}. 
}
\label{fig:ns}
\end{figure}

In this work, we propose a parameter block-wise truncation scheme (``correlated" truncation) based on high-dimensional ratio estimators that accounts for correlations between parameters. This correlated truncation region is defined through a hard likelihood-to-evidence ratio constraint,
\begin{equation} \label{eq:block}
    \tilde p_R(\boldsymbol \vartheta) = \frac1Z \mathbb{I}(\boldsymbol \vartheta \in \Gamma_{\boldsymbol \vartheta}^{(R-1)}) p(\boldsymbol \vartheta) \,,
\end{equation}
where $\Gamma_{\boldsymbol{\vartheta}}^{(R - 1)}$ is as defined in Eq.~\eqref{eq:gamma_r}. In Fig.~\ref{fig:ns}, we show a visual comparison of the two prior truncation strategies in a simple 2-dimensional parameter space. However, following this approach brings about a new challenge that will be addressed in the next section: how does one efficiently define the boundaries of $\Gamma_{\boldsymbol \vartheta}^{(R)}$ and sample from within it?

\subsection{Sampling from ratio estimators: posterior and constrained prior samples} \label{subsec:ns}

\noindent As discussed in Sec.~\ref{sec:intro}, NRE does not have sampling functionalities and Monte Carlo-based sampling algorithms are used to obtain samples from the approximate posterior.\footnote{It is worth noting that this is somewhat contrary to NPE, where one can directly draw posterior samples from the normalising flow.} Moreover, we are not only interested in posterior samples, but also in how we can efficiently draw constrained prior samples from a region defined through a hard likelihood-to-evidence ratio constraint. In this work, we propose utilizing a \emph{slice-based nested sampling algorithm}, as an efficient method to sample both posterior samples and constrained prior samples from the ratio estimator. Importantly, constrained prior samples can be obtained through the same nested sampling techniques. Our sampler choice is strongly inspired by the success of slice-based nested samplers in efficiently and reliably exploring high dimensional parameter spaces \citep{Handley_2015_polychord, Handley_2015_cosmo}. Additionally, slice sampling was proposed in \cite{papamakarios2019sequential} as a sampler for NLE. 

First proposed in \cite{Skilling_2006_ns}, nested sampling allows one to sample from high-dimensional probability densities by evolving an ensemble of live points through a high-dimensional parameter space. At the core of nested sampling algorithms is the problem of constrained prior sampling within iso-likelihood contours \citep{Ashton_2022}. This opens up the possibility to re-use technology developed for nested sampling for the purpose of sampling not only from the posterior, but also from the constrained prior region, as defined in Eq.~\eqref{eq:gamma_r} within iso-likelihood-to-evidence ratio contours.\footnote{In traditional implementations the contours are defined by iso-likelihood levels, not iso-likelihood-to-evidence ratio levels.}

Typically, nested sampling algorithms start by drawing a collection of live points from the prior. In the current context, one evolves them by discarding the point with the lowest ratio, denoted with $r_\mathrm{min}$, and replacing it with a new point subject to the constrain $r > r_\mathrm{min}$. The remaining live points are now uniformly distributed over a compressed volume (as they are drawn from a constrained prior as in Eq~\eqref{eq:gamma_r} with $\epsilon = r_\mathrm{min}$).

The most challenging aspect of the nested sampling algorithm is drawing new live points under the hard likelihood-to-evidence ratio condition $r > r_\mathrm{min}$. One possible reliable and efficient way to do so is through slice sampling, first introduced in \cite{Neal_2000}. Starting from one randomly chosen live point, slice sampling builds a chain of proposed live points by taking sequential 1-dimensional steps in a random direction. The length of this chain controls the amount of correlation between the new live point and the initial one (for further information see \cite{buchner2023nested}).

In our slice sampling implementation, we use a number of different ploys to make our sampling scheme more efficient for the purposes at hand. Firstly, instead of a single chain, the user can define the number of slice sampling chains to draw new live points that will be run in parallel. Secondly, the sampler is implemented such that it allows vectorized evaluations of the natively GPU-based ratio estimator, which provides considerable computational speed-up. We also provide useful functionality to compute the threshold $\epsilon$ that defines the constrained prior region in Eq.~\eqref{eq:gamma_r}, based on how much probability mass from the current approximate likelihood-to-evidence ratio one wants the region to include. This is relevant for setting the convergence criterion related to posterior mass that defines the iso-likelihood-to-ratio contour, as in Eq.~\eqref{eq:gamma_r}. Importantly, inference errors caused by truncation arise when not enough posterior mass is included in the truncated regions, resulting in wrongly excluded parts of the parameter space of interest (for a detailed discussion and test of the impact that overly tight complex truncation induces on inference, we refer the reader to App. 6.10 of \citealt{deistler2022truncated}). 
For sequential SBI applications, we advise the user to conservatively choose the threshold $\epsilon$ such that the truncated prior encloses the highest probability density (HPD) region that contains at minimum $99.9 \%$ (\textit{i.e.}~$1 - 10^{-3}$) of the probability mass. 

\paragraph*{Network stability during sampling.}
During the sampling step, numerical issues can arise because the slice sampler may evaluate the network outside of the current truncation region $\Gamma^{(R)}_{\boldsymbol{\theta}}$ when searching for new points.
In order to detail the issue, we recall our definition of $B={p(\boldsymbol \theta)}/{\prod_{i=1}^dp(\theta_i)}$, which is zero outside of the support of the joined distribution $p(\boldsymbol \theta)$. Formally, this is not a problem for the ratio $A/B$ (see Eq.~\ref{eq:AB}), because that region coincides with the prior region that we truncated away in our truncation procedure (\textit{i.e.} regions of very low posterior mass). In practice, during sampling, since the network in a given round $R$ has not seen training samples outside of the constrained prior $\tilde p_R(\boldsymbol \vartheta)$ (see Eq.~\ref{eq:block}), this can lead to estimated values of our neural likelihood-to-evidence ratio $A/B$ which are much larger than the formal expectation (indeed $B = 0$ exactly outside this region).

We solve this stability issue with the following strategy, that has no impact within the truncation bounds $\Gamma^{(R)}_{\boldsymbol{\theta}}$, and therefore on the results.
During the sampling step, the following substitution is performed in the network
\begin{equation}
    B \rightarrow \text{max}(B, \min_{\Gamma^{(R)}_{\boldsymbol{\theta}}}B). 
\end{equation}
Hence, $B$ can be at minimum as small as the smallest value it can take inside the truncation bound $\Gamma^{(R)}_{\boldsymbol{\theta}}$. As a result, the substitution will never happen inside the constrained region $\Gamma^{(R)}_{\boldsymbol{\theta}}$ where the network was trained, and consequently will not have any impact on the new truncation set $\Gamma^{(R+1)}_{\boldsymbol{\theta}} \in \Gamma^{(R)}_{\boldsymbol{\theta}}$. We further emphasize that by construction, the truncation region should never expand throughout a round, so this new substitution only plays the role of fixing the stability issues outside the constrained prior.

\vspace{10pt}
\noindent To summarise, in this section we have presented the main contributions of this work in the form of two new building blocks of the TMNRE framework: autoregressive NRE and a slice-based nested sampler to sample posterior and constrained prior samples from a ratio estimator. In the next section, we will apply these newly developed tools to a toy Gaussian example, as well as two astrophysical applications -- stellar streams and strong lensing.

\section{Experiments} \label{sec:experiments}

\noindent We first apply our newly developed tools to a toy multivariate Gaussian example. For more realistic settings, we consider two astrophysics problems: stellar streams parameter inference and substructure parameter inference in strong gravitational lensing observations.

\subsection{Multivariate Gaussian: scalability} \label{toy}

\begin{figure}
\includegraphics[width=\linewidth]{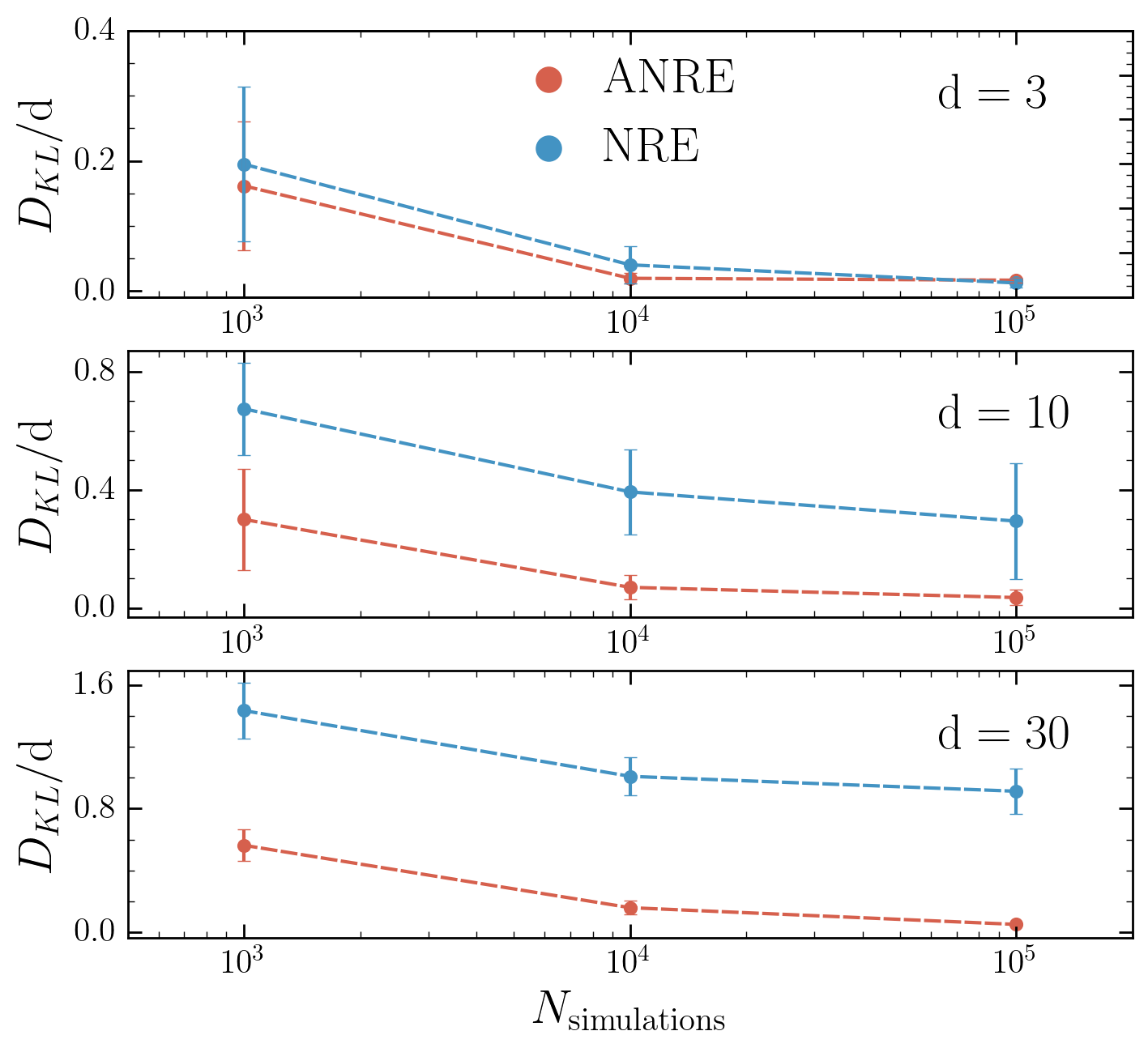}
\caption{
 Multivariate Gaussian toy example results. In the three panels we show the KL divergence value between the analytical posterior distribution and the estimated ones from the autoregressive (ANRE, in red) a non-autoregressive (NRE, in blue) models. The metric is shown as a function of simulation budget in three different dimensionalities ($d=3, 10, 30$). For our fixed network capacity, NRE is able to perform as well as ANRE for high simulation budget in low-dimensional scenarios. For higher-dimensionality, the quality of posterior samples obtained from the autoregressive model matches significantly more closely the one from the analytic solution based on the KL divergence value.}
\label{fig:toy}
\end{figure}

\noindent In this first application, we are interested in testing how the performance of ``vanilla" NRE and autoregressive NRE models scale with dimensionality and simulation budget. To do so, we consider a $d$-dimensional Gaussian toy model with strong correlations. The parameters $\boldsymbol \vartheta \in \mathbb{R}^d$ are drawn from a multivariate normal prior $p(\boldsymbol \vartheta)=\mathcal{N}(\mathbf 0, \mathbf{\Sigma_0})$, with diagonal covariance matrix $\mathbf{\Sigma_0}= 0.1 \odot \mathbf I$. The observations $\mathbf x \in \mathbb{R}^d$ are drawn from a multivariate normal likelihood $p(\mathbf x \mid \boldsymbol \vartheta)= \mathcal{N}(\boldsymbol \vartheta, \mathbf{\Sigma})$, with fixed covariance $\mathbf{\Sigma}$, which has correlation scales of $0.1$ for the off-diagonal entries, and $0.11$ for the diagonal ones. As a reference, we use the analytic solution for the true posterior given by Bayes' theorem.

We test NRE and autoregressive NRE for dimensions $d = 3, 10, 30$ and for different simulation budgets of $N_\mathrm{simulations} =10^3, 10^4, 10^5$ training data examples. Each conditional ratio in the autoregressive network is modelled with $4$ ResNet \citep{he2015resnet} blocks with $64$ hidden features (implemented in \texttt{swyft}). The ``vanilla" NRE network is instead modelled with a single network whose number of ResNet blocks and parameters is adjusted accordingly to match the total number of parameters in the autoregressive model. 

To compare the performance of the two models we consider the Kullback-Leibler (KL) divergence $D_\mathrm{KL}$ between the posterior estimated by each model and the analytical solution. The KL divergence is a statistical distance that measure the difference between two probability distributions; the closer it is to zero, the better agreement there is. The mean and the covariance of the estimated posterior are computed from the posterior samples of the two models, obtained with our slice-based nested sampler. 
Our results are shown in Fig.~\ref{fig:toy}, where each point is the mean KL divergence of 10 different observations. The error bar represents the standard deviation of these values. In low-dimensional scenarios ($d=3$), NRE is able to perform as well as autoregressive NRE for high simulation budget. For higher-dimensionality, the quality of posterior samples obtained from the autoregressive model matches significantly more closely the one from the analytic solution for lower simulation budgets. In App.~\ref{apx:toy}, we show the comparison between the analytical posterior and the estimated posterior with simulation budget $N_\mathrm{simulations} = 10^{5}$ from the two models for one observation with $d=10$.

\subsection{Stellar streams: autoregressive variable ordering} \label{subsec:stream}

\begin{figure*}
    \centering
     \begin{tikzpicture}[every node/.style={anchor=center}]
        \node(a) at (8,4){ \includegraphics[width=\linewidth]{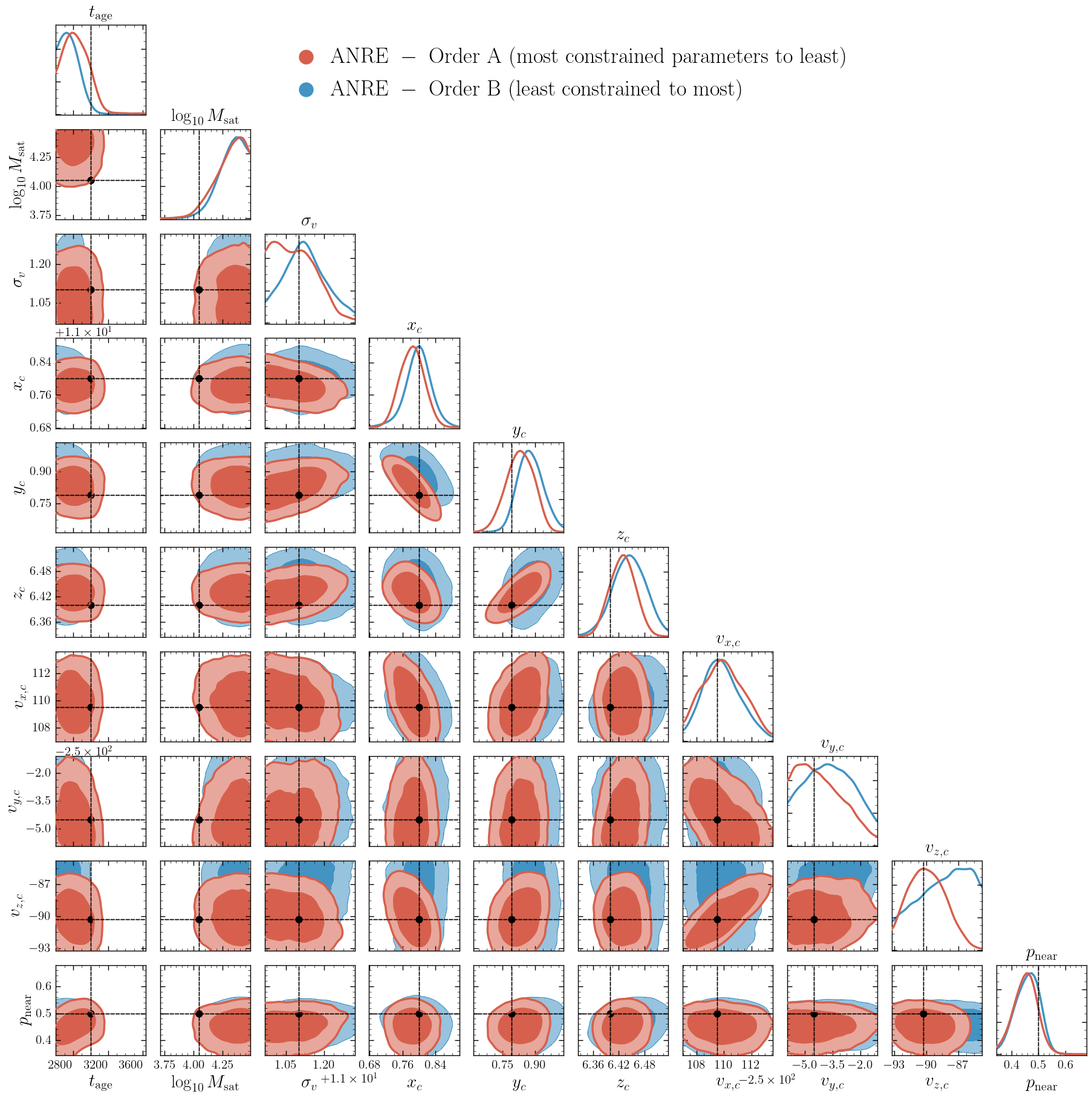}};
        \node(b) at (13.5,7){\includegraphics[height=7cm]{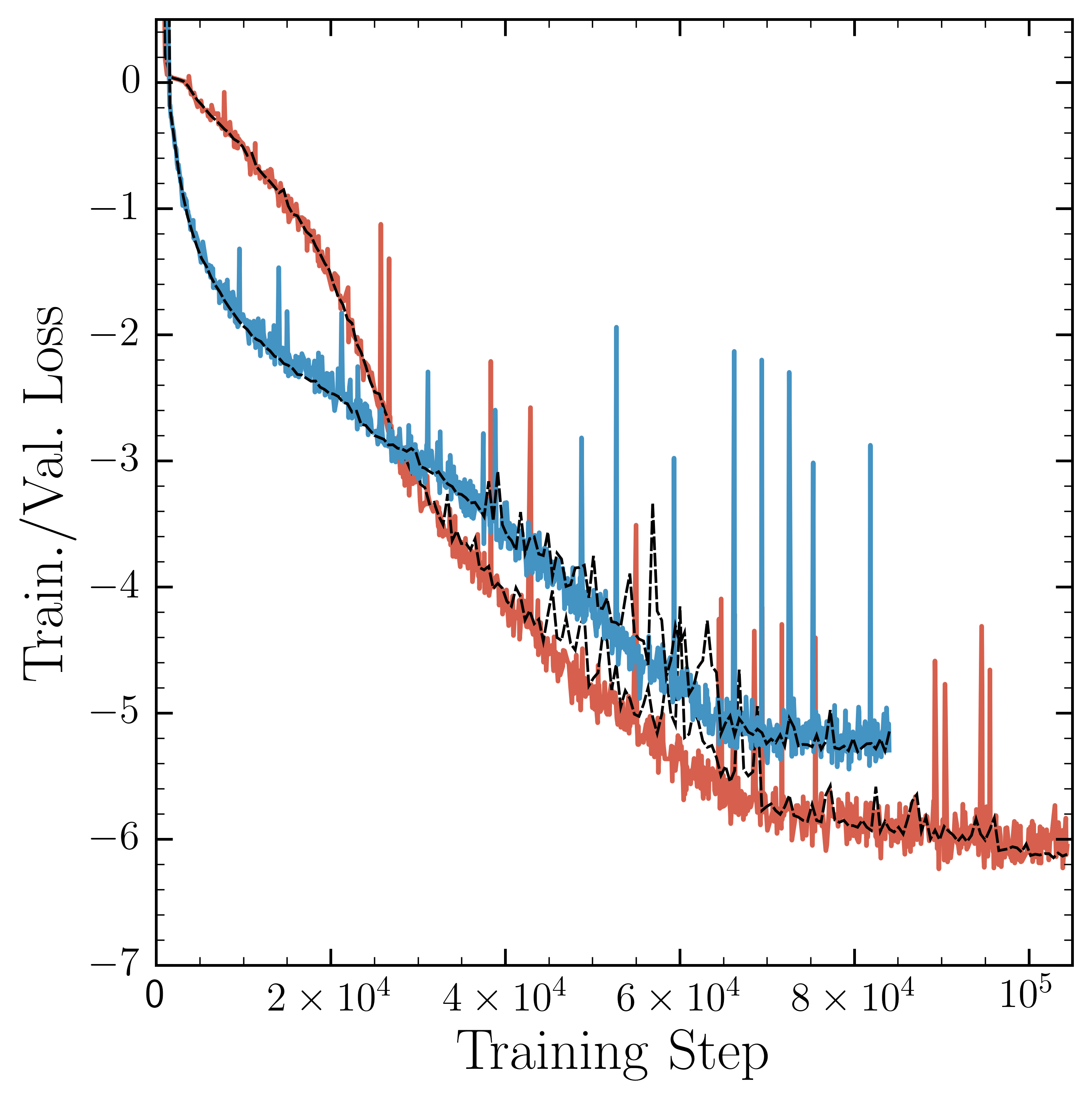}};
    \end{tikzpicture}
    \caption{Stellar streams experiment results. \emph{Corner plot:} The corner plot shows the comparison between the inference results using different parameter ordering schemes across those parameters that are constrained in the analysis (for the full set, see App.~\ref{apx:streams}). Order A (most constrained parameters to least constrained) is shown with red contours, whilst Order B (least constrained to most) is shown in blue. We see that Order A performs marginally better in terms of precision, although the improvement is somewhat marginal. The true injected values are shown with black dashed lines and dots. \emph{Upper right inset:} This inset shows the training loss (see Eq.~\eqref{eq:loss}) as a function of the number of training steps for the two orderings. We see that Order A achieves a lower value of the loss, in line with the conclusions from the corner plot. The validation loss for each case is shown with black dashed lines.}
    \label{fig:streams_loss}
\end{figure*}

\noindent Stellar streams are very old, dynamical objects in the Milky Way that form as the result of the tidal disruption of objects such as globular clusters or small dwarf galaxies as they orbit the host. In principle, they can act as detailed tracers of the galactic potential, the disruption history and physics of the star clusters, and even possible dark substructures in the Milky Way (see \textit{e.g.} \cite{Erkal:2015kqa,Banik:2018pjp,Banik:2018pal,Koposov:2009hn,Bonaca:2014qia,Amorisco:2016evb,Bovy:2016chl,Malhan:2022nfe} for examples of the existing analyses). On the other hand, robust analysis of these objects is challenging for a number of reasons. Firstly, simulating the stream in full generality is computationally challenging, so delicate modelling choices need to be made. Secondly, the statistical properties of the resulting stream are difficult to write down explicitly in the form of a likelihood, mainly due to complicated observational effects such as membership probabilities in \textit{e.g.} Gaia data \citep{Gaia:2018aaa,Gaia:2021aaa}. Combined with the general simulation efficiency that has been observed with neural ratio estimation, this strongly motivates the application of SBI techniques to such a problem, something that has been studied in detail in \cite{alvey2023albatross} (see also \cite{Hermans:2020skz}).

In the current context, we will use the inference of stream model parameters in an identical set up to \cite{alvey2023albatross} to highlight a relevant aspect of the autoregressive algorithm presented in this work. Specifically, we use the case study to highlight the implications that varying the ordering of parameters in the conditional distributions product found in Eq.~\eqref{eq:A} has on inference performance. The intuition that we wish to test is the following: we expect that the distribution $p(\vartheta_i | \mathbf x)$ is most useful to learn if $\vartheta_i$ is well constrained. Similarly, if $\vartheta_j$ is not well constrained, then adding conditional information \textit{e.g.} in $p(\vartheta_j | \mathbf{x}, \boldsymbol \vartheta_{1:j-1})$ can add relevant information. With this said, the expectation is that an approximately optimal ordering runs from $\vartheta_1$ as the most constrained parameter (relative to the prior), and $\vartheta_d$ as the least constrained parameter.

Given the analysis in \cite{alvey2023albatross}, we have good expectations about the relative constraining power of the streams analysis on various parameters. As such, we choose two orderings: Order A which runs from most to least constrained, and Order B which runs vice versa (see App.~\ref{apx:streams} for details). We train an autoregressive estimator in both cases to perform inference on all parameters of the model. The results of this experiment are shown in Fig.~\ref{fig:streams_loss} in terms of the resulting posterior distributions, as well as the training and validation loss curves for the neural network training process. 

There are two things to note from Fig.~\ref{fig:streams_loss}: firstly, the ordering of parameters does have an effect on the training procedure, with the expectation laid out above clearly realised. In other words, we see a plateauing of the optimal training and validation loss for Order B at a value that is clearly larger than the asymptotic value in the case of Order A. In addition, broadly the overall quality of posterior inference on parts of the model that are well measured is of higher quality with Order A. Together, this suggests that the general intuition regarding parameter ordering in autoregressive models holds here also. With that being said, however, we do observe that the `penalty' for choosing a non-optimal ordering is not severe, at least in this case, and high fidelity inference could be achieved by \textit{e.g.} increasing the simulation budget (see App.~\ref{apx:streams} for an example), or following the prior truncation scheme described above and iterating the simulation generation, training, and inference steps. To be clear on this last point, this example is designed to demonstrate for fixed simulation budget and fixed network architecture the impact of variable ordering. We expect in this particular case that if the prior truncation scheme is followed, the posteriors will continue to shrink and converge to the same answer. For high precision inference, we advise that on convergence, an explicit test varying the ordering is carried out to confirm that the posteriors do not shift by more than the desired sensitivity. For full technical details regarding this example, see \cite{alvey2023albatross} or App.~\ref{apx:streams}.  Also in App.~\ref{apx:streams}, we show two other random orderings to highlight the relative insensitivity to this choice and build on the example in Sec.~\ref{toy}.

\begin{figure*}
 \begin{tikzpicture}[every node/.style={anchor=center}]
    \node(a) at (8,4){ \includegraphics[width=\linewidth]{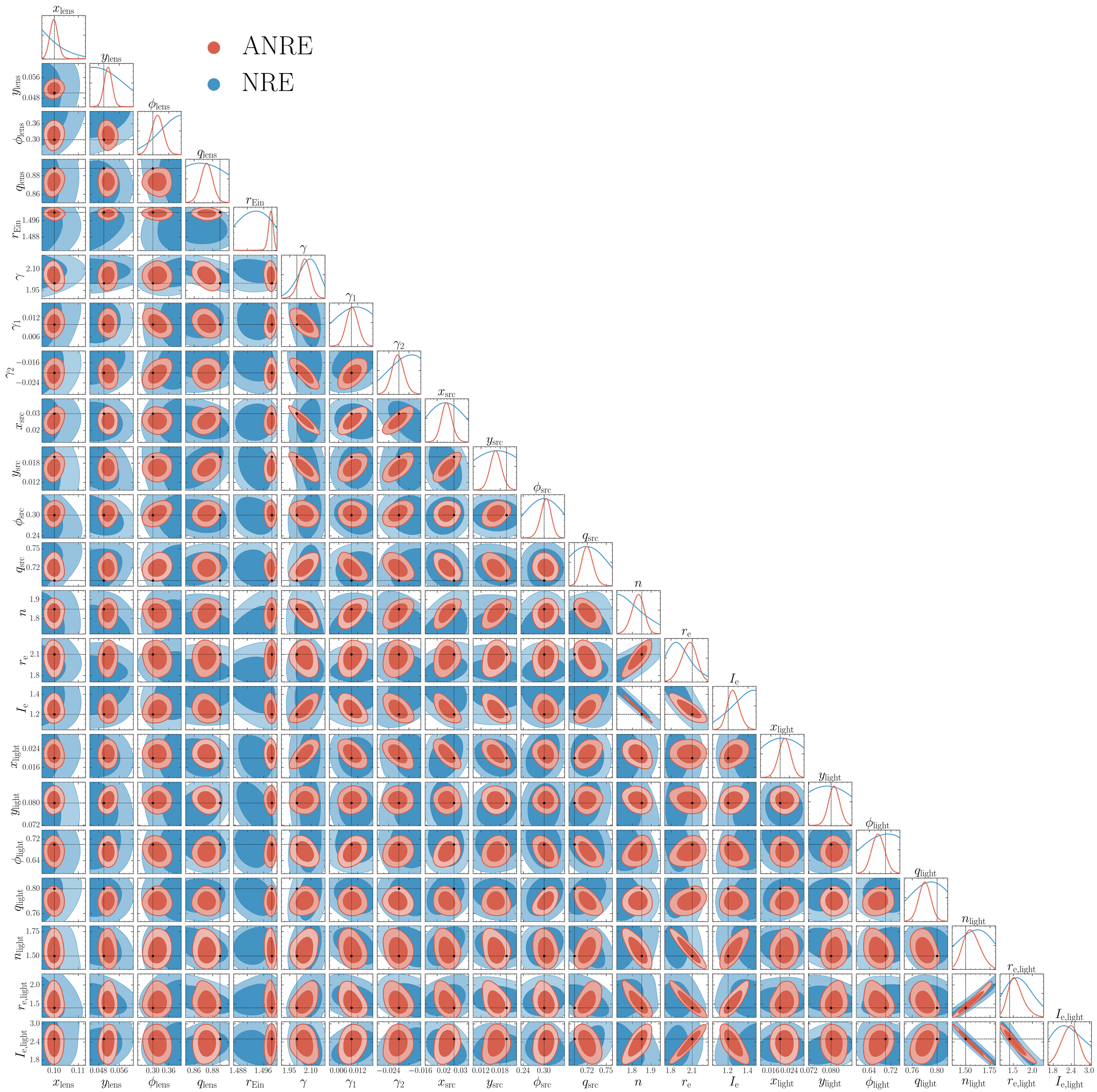}};
    \node(b) at (12,9){\includegraphics[height=5.5cm]{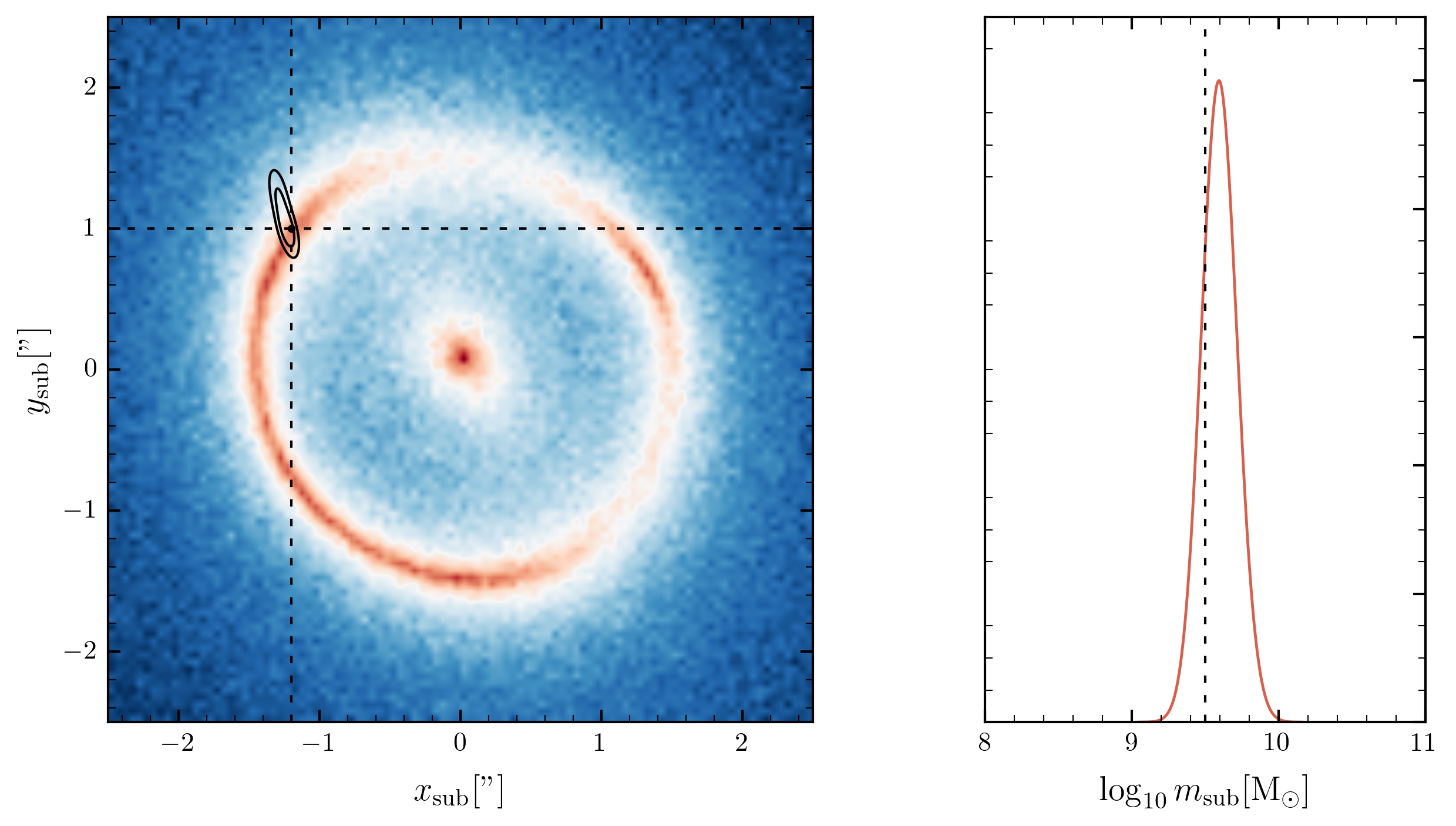}};
\end{tikzpicture}
\caption{Strong gravitational lensing experiment results. \emph{Corner plot:} The corner plot shows the comparison between the macro-model parameter inference results using the autoregressive NRE model (ANRE, red contours) and the non-autoregressive one (NRE, blue contours). The true values are shown with black dashed lines and dots and are shown in Tab.~\ref{tab:lensing-params}. \emph{Upper right inset:} The inset shows the results for subhalo parameter inference for our target mock observation.
}
\label{fig:corner}
\end{figure*}

\subsection{Strong gravitational lensing: correlated truncation} \label{subsec:lensing}

\noindent Galaxy-galaxy strong gravitational lensing occurs when the paths of light rays from a background galaxy are distorted by the mass of an intervening lens galaxy before entering a telescope \citep{lensing}. This leads to extremely distorted ring-shaped images with multiple copies of the source. If a small-scale dark matter halo is present in such a system, its distortions will be localised, mostly impacting the image of the source along the nearest line of sight. As first proposed in \cite{Mao_1997}, by carefully analyzing the relationship between the multiple images of the source, the distortions due to a dark matter perturber can be disentangled from possible variations in the source light and its properties can be measured.

In reality, substructure parameter inference in strong gravitational lensing systems is a very challenging task for a number of reasons. In particular, the desired signal manifests as percent level variations, lensing systems are remarkably complicated, and the variance between images is high (\textit{i.e.} observations are very diverse). A targeted SBI approach to this ``needle in a haystack" type of analysis is therefore well-motivated and has been proven successful, see \textit{e.g.} \cite{coogan2020targeted, coogan2022walks}. 

For this application, we consider a lensing system with the following components: source light, lens light, lens mass distribution, and external shear (which we collectively call the macro-model); and one subhalo. More details on the system and noise model can be found in App.~\ref{apx:lensing}. The blending effect due to the overlap of the light emitted by the lensed source and by the lens itself complicates the interpretation and analysis of observed lensed systems, and makes it challenging to isolate and study the properties of the lensed source. This blending effect can be mitigated if multi-wavelength observations are available. Usually the lens light gets subtracted assuming the best-fit value (see \textit{e.g.} \cite{Vegetti_2012}). In this SBI application, we leave it free to vary and infer its parameters at the same time as all other components, accounting in the analysis for its uncertainties and the correlations it has with the rest of the system components.

\vspace{10pt}
We perform inference on the simulated target observation shown in Fig.~\ref{fig:corner}, generated with the parameters in Tab.~\ref{tab:lensing-params}. The first step in the analysis is to reduce training data variance by constraining the macro-model parameters prior. Starting from the full macro-model prior shown in Tab.~\ref{tab:lensing-params}, we use the procedure described in \cite{coogan2022walks, Anau_Montel_2022} to constrain it sequentially via TMNRE rounds using box truncation. When convergence\footnote{We define the ratio estimator to be converged when, after two consecutive rounds in which we double up the simulation budget, the constrained hyper-rectangular box prior has shrunk by less than 10\%.} is reached, we are not able to reduce any further the data variance displayed by the simulations via 1-dimensional marginal posterior estimation and box truncation.

\begin{figure}
\centering
\begin{minipage}{\linewidth}
\begin{tikzpicture}[baseline=(current bounding box.center), scale=0.75]

  \definecolor{myblue}{RGB}{66, 117, 176}
  \definecolor{myred}{RGB}{174, 60, 60}
  
  \draw[very thick] (-5, -5) rectangle (0, 0);

  \pgfmathsetmacro{\startx}{-5}
  \pgfmathsetmacro{\endx}{-0.2}
  \pgfmathsetmacro{\numx}{24}
  \pgfmathsetmacro{\starty}{-0.2}
  \pgfmathsetmacro{\endy}{-5}
  \pgfmathsetmacro{\numy}{24}
  
  \foreach \ix in {0,...,\numx} {
    \foreach \iy in {0,...,\numy} {
        \ifthenelse{\equal{\ix}{\iy}}{
        \pgfmathsetmacro{\x}{\startx + \ix * (\endx - \startx) / \numx}
        \pgfmathsetmacro{\y}{\starty + \iy * (\endy - \starty) / \numy}
        \draw [fill=myblue, opacity=0.7] (\x, \y) rectangle ++(0.2, 0.2);
        }{}
    }
   }

   \draw[thick,decorate,decoration={calligraphic brace,raise=1ex}]
    (-5,0) -- node[above right=0.6ex,align=center,font=\sffamily]{$\boldsymbol \vartheta_\mathrm{macro}$}(-0.6,-4.4);
    \draw[thick,decorate,decoration={calligraphic brace,raise=1ex, mirror}]
    (-0.6,-4.4) -- node[left=1ex,align=center,font=\sffamily]{$\boldsymbol \vartheta_\mathrm{sub}$}(0,-5);

  \draw[very thick] (1, -5) rectangle (6, 0);

  \draw [fill=myblue, opacity=0.7] (1, -4.4) rectangle ++(4.4, 4.4) node [midway, align=center,font=\sffamily]{$\boldsymbol \vartheta_\mathrm{macro}$}; 
  \draw [fill=myblue, opacity=0.7] (5.4, -5) rectangle ++(0.6, 0.6);
  \draw (4.7,-4.7) node {$\boldsymbol \vartheta_\mathrm{sub} \rightarrow$};

\end{tikzpicture}
\end{minipage}
\caption{Prior truncation strategies. \emph{Left:} With a parameter-wise truncation strategy, each parameter gets individually constrained with boxes. \emph{Right:} Using a parameter block-wise truncation strategy, we divide the parameter space in blocks (lensing macro-model parameters $\boldsymbol \vartheta_\mathrm{macro}$ and subhalo parameters $\boldsymbol \vartheta_\mathrm{sub}$), depending on what dominates the data variance, and which parameters are most correlated. During sequential inference, we account for correlations between parameters in the same block through correlated truncation.}
\label{fig:block}
\end{figure}
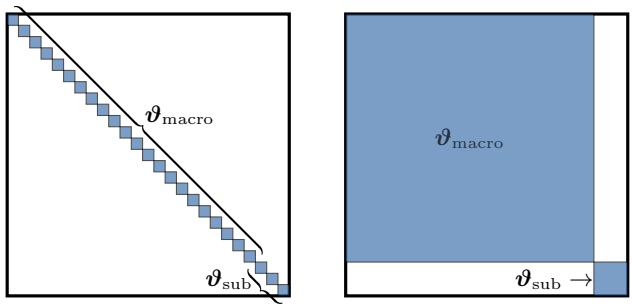

As previously discussed in Sec.~\ref{subsec:truncation}, a disadvantage of using hyper-rectangular boxes lies in the fact that for high-dimensional and correlated parameter spaces the constrained region contains more probability mass than required. We will now employ a parameter block-wise correlated prior truncation strategy instead of box truncation, as presented in Sec.~\ref{subsec:truncation}. In Fig.~\ref{fig:block}, we show a simple visualization of the two prior truncation strategies for this specific application.

In order to learn the macro-model parameter correlations, especially the ones between lens light and source light, we model the joint posterior of the  macro-model parameters. We train two models to estimate the macro-model joint ratio: a ``vanilla" NRE model and the autoregressive NRE model presented in Sec.~\ref{subsec:anre}. We used the same simulation budget and amount of network weights (for more details regarding training and the employed neural network architecture, see App.~\ref{apx:lensing}). In Fig.~\ref{fig:corner}, we compare the posterior samples obtained from the two models using our slice-based nested sampler presented in Sec.~\ref{subsec:ns}. It is clear that, for this application and simulation budget, autoregressive NRE performs significantly better than NRE, which is not able to properly model the joint macro-model parameter distribution.

\begin{figure}
\centering
\begin{minipage}{\linewidth}
\begin{tikzpicture}
  \node (img)  {\includegraphics[width=0.8\linewidth]{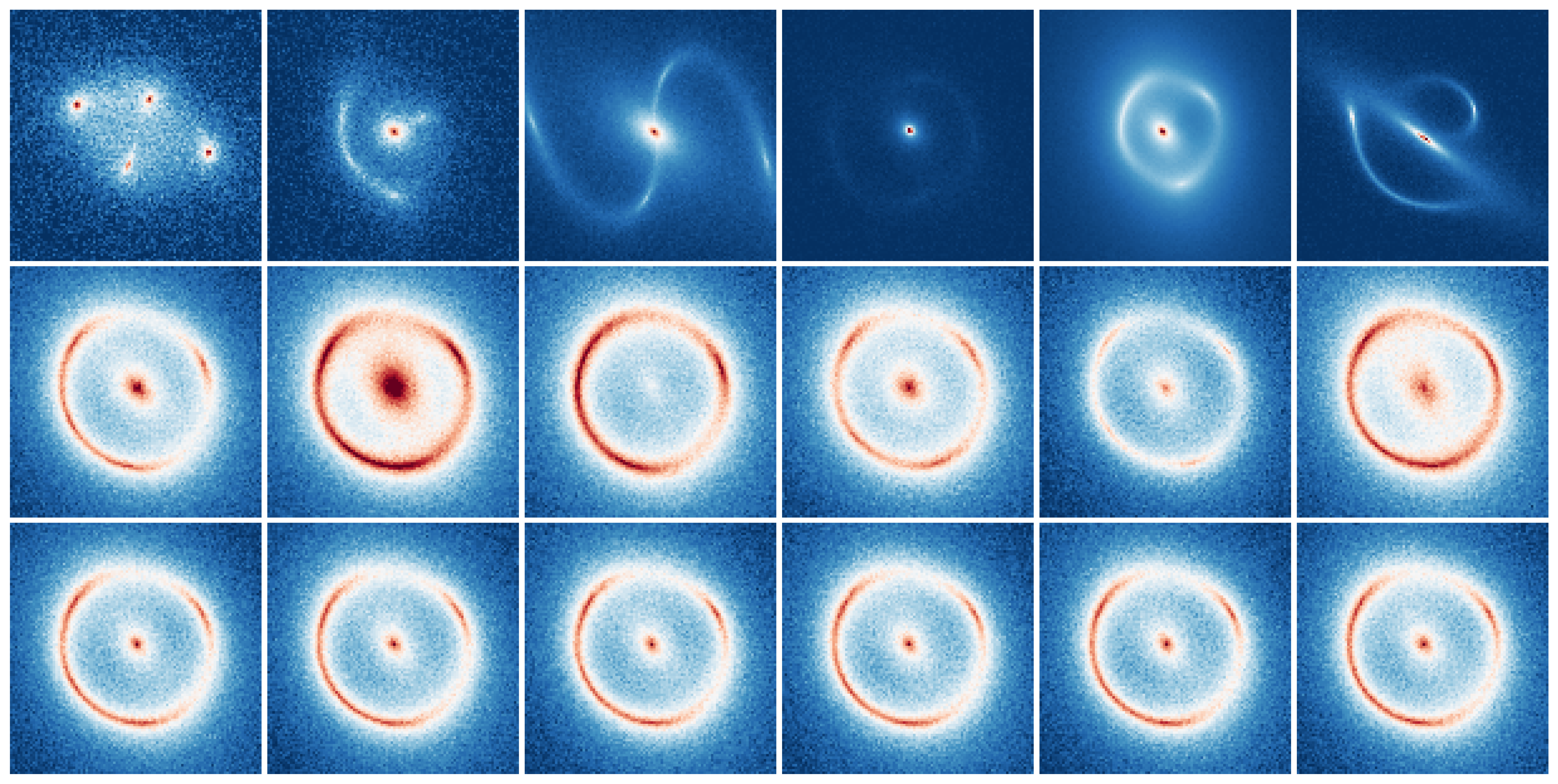}};
  \node[left=of img, node distance=0cm, text width=1cm, yshift=+1.1cm, anchor=center] {Full prior};
  \node[left=of img, node distance=0cm, text width=1cm, anchor=center] {Box\\ truncation};
  \node[left=of img, node distance=0cm, text width=1cm, yshift=-1.1cm, anchor=center] {Correlated\\ truncation};
 \end{tikzpicture}
\end{minipage}%
\caption{Targeted training data. The simulations in the first row are drawn from the full initial prior, described in Tab.~\ref{tab:lensing-params}. In the intermediate row, we show simulations used for the last round of training of the 1-dimensional marginal ratio estimators, where the macro-model parameters were drawn from the most constrained hyper-rectangular box prior. In the last row, the macro-model parameters were sampled from the high-dimensional constrained prior with the slice-based nested sampler. The images in the second and last row are plotted in the same color scale, whereas the ones in the first row are allowed to vary for visualization purposes since they exhibit a larger variance.
}
\label{fig:targeted_data}
\end{figure}
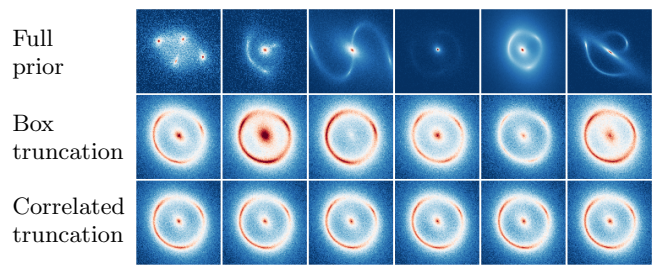

We then sample correlated constrained prior samples with our slice-based nested sampler. By modeling and exploiting the intricate correlations between lens light, source light, and lens parameters, instead of using a parameter-wise truncation strategy, we are able to achieve a much lower data variance. In Fig.~\ref{fig:targeted_data}, we display examples of training data. The simulations in the first row are drawn from the full initial prior. In the intermediate row, we used macro-model parameters drawn using the box truncation scheme. Finally, the simulations presented in the last row use the slice-based nested sampler to sample the macro-model parameters from the high-dimensional constrained prior, using the autoregressive model.

Having reduced the macro-model parameters training data variance, we are then able to focus on substructure parameter inference, for which the results are shown in Fig.~\ref{fig:corner}. These results were obtained with two sequential rounds of inference using a correlated truncation scheme, as explained in Sec.~\ref{subsec:sequential}. In contrast, given the initial fixed simulation budget, we were not able to constrain the subhalo parameter in our target observation with the data variance displayed by simulations using only box truncation.

\section{Discussion and conclusions} \label{sec:conclusion}

\noindent The motivation for the SBI method development presented in this work comes from the data analysis and statistics challenges facing the fields of astrophysics and cosmology in light of high quality data from current and future experiments \citep{laureijs2011euclid,JWST2006,lsst,elt,gaia,SKA,CTA}. In this work, we have focused on one specific SBI algorithm, TMNRE, and advanced its potential with the introduction of two new components:
\begin{enumerate}[a)]
    \item \emph{Autoregressive NRE}, presented in Sec.~\ref{subsec:anre}, estimates multi-dimensional (marginal) posteriors by effectively computing $d$ 1-dimensional conditional ratios instead of a $d$-dimensional joint one (see Eq.~\eqref{eq:autoregressive}). This way of approaching high-dimensional distribution estimation through an autoregressive scheme has been motivated by its proven scaling potential in other SBI techniques \citep{Germain_2015, Uria_2016, Papamakarios_2018_masked}. We release a publicly available implementation of the autoregressive model in the TMNRE package \texttt{swyft}\footnote{\url{https://github.com/undark-lab/swyft}}.
    \item Our \emph{slice-based nested sampling} implementation, presented in Sec.~\ref{subsec:ns}, offers vectorized evaluations of the natively GPU-based ratio estimator and introduces a convergence criterion related to posterior mass. The latter is essential for sequential inference applications since it allows one to define the truncation region in Eq.~\eqref{eq:gamma_r}. The sampler choice naturally comes from the realisation that nested sampling techniques provide a tool to efficiently sample not only from a multi-dimensional posterior, but also natively from a constrained prior. Our slice sampler implementation is available in the package \texttt{torchns}\footnote{\url{https://github.com/undark-lab/torchns}}.
\end{enumerate}
We have demonstrated their application in three case studies presented in Sec.~\ref{sec:experiments}. Firstly, we tested the performance of the autoregressive ratio model against the non-autoregressive one in a toy example. The main results, for different simulation budgets and dimensionality, are presented in Fig.~\ref{fig:toy}. Secondly, we explored how variable ordering impacts the autoregressive model in a stellar streams analysis. We have found that a non-optimal variable ordering does have an impact, but does not severely penalise the overall inference result in our specific application, as shown in Fig.~\ref{fig:streams_loss}. Lastly, we investigated the potential of a correlated truncation scheme in a proof-of-concept application to substructure searches in a strong gravitational lensing image. The main outcomes are exhibited in Figs.~\ref{fig:corner} and~\ref{fig:targeted_data}: in particular, given our fixed simulation budget, the non-autoregressive model is not able to estimate the macro-model parameter joint distribution, and only through correlated prior truncation are we able to constrain the subhalo parameters.

\vspace{10pt}
\noindent \emph{Outlook.} There are two aspects to consider as far as outlook is concerned for our method development. The first concerns the known limitations of the autoregressive modelling and the nested sampling techniques. The second is more general, and considers the analysis settings in which this approach may be useful.

As discussed above, one of the potential limitations of autoregressive modelling is its sensitivity to parameter orderings. Although we investigated this in the context of stellar streams in Sec.~\ref{subsec:stream}, and found that the effect was minimal, it is possible that this is model dependent. As such, to extend the method, it would be interesting to develop techniques to derive an (approximately) optimal ordering of the parameters, \textit{e.g.} by estimating the strength of correlations between various parameter sets. 

On the nested sampling side, known limitations include the fact that it becomes very costly in terms of the required number of network evaluations for high dimensional parameter spaces, say $d \gtrsim 30$ (for a discussion see, \textit{e.g.}, \cite{buchner2023nested}). As such, it will be interesting to further investigate methods that exploit gradient information and scale better to high dimensions, such as the proximal nested sampling technique proposed in \cite{Cai2021ProximalNS}. Moreover, our current implementation of the slice sampler does not include clustering detectors (see \textit{e.g.} \cite{Feroz_2008_multimodal_ns}), which makes it inefficient for multi-modal posteriors.

More generally, we believe the method we have developed could be useful in the following type of scenario: suppose that a part of the parameter space dominates the data variance, but is not the one that we are ultimately most interested in performing inference on. Using traditional methods, there are typically two approaches: a) solve the full joint analysis problem for all parameters, but face significant computational challenges, or b) analyse the nuisance parameters separately, and then fix them to some form of best-fit value to use in the rest of the analysis, at the cost of neglecting their uncertainties. For example, as briefly mentioned in Sec.~\ref{subsec:lensing}, this is exactly the scenario in standard strong gravitational lensing analyses where the lens light gets subtracted assuming the best-fit value, before analysing the lensed emission.

With the method developed in this work, however, there is an alternative approach available. In particular, as mentioned in Sec.~\ref{subsec:sequential}, the fact that prior truncation composes well with marginalisation allows us to consistently combine different analysis strategies that target distinct parts, or ``blocks", of the model into a coherent ``inference assembly". This exact procedure was exemplified in Sec.~\ref{subsec:lensing}, where we first reduced training data variance by truncating the lens and lensed source parameters, and then performed sequential inference on our parameters of interest, the ones pertaining the substructure.

In conclusion, this approach can be relevant for many different cosmological and astrophysical applications, where the analysis would be carried out simultaneously on different components of the systems, \textit{e.g.}~parameters of interest, foregrounds, backgrounds, nuisance parameters, instrumentation parameters \textit{etc}. Ultimately, this will allow us to flexibly combine different inference strategies in order to draw coherent and consistent conclusions based on the full model and all the data.

\section*{Acknowledgements}

We thank Will Handley and Kilian Scheutwinkel for useful discussions. This work is part of a project that has received funding from the European Research Council (ERC) under the European Union’s Horizon 2020 research and innovation program (Grant agreement No. 864035 -- UnDark). JA is supported through the research program ``The Hidden Universe of Weakly Interacting Particles" with project number 680.92.18.03 (NWO Vrije Programma), which is partly financed by the Nederlandse Organisatie voor Wetenschappelijk Onderzoek (Dutch Research Council). 
This work was carried out on the Snellius Compute Cluster at SURFsara. We acknowledge the use of the \texttt{python} \citep{python} modules, \texttt{matplotlib} \citep{matplotlib}, \texttt{numpy} \citep{numpy},  \texttt{scipy} \citep{scipy}, \texttt{AstroPy} \citep{astropy}, \texttt{PyTorch} \citep{pytorch}, \texttt{tqdm} \citep{tqdm}, and \texttt{jupyter} \citep{jupyter}.

\bibliography{references} 
\newpage
\clearpage

\appendix

\section{Another formulation} \label{apx:anre}

\noindent The autoregressive model presented in Sec.~\ref{subsec:anre} can be also written with a single network that can estimate both components. In order to do this, we introduce an auxiliary variable $c = -1, 1$.  We then consider the ratio
\begin{equation}
    r(\vartheta_i; \mathbf x, c, \boldsymbol \vartheta_{1:i-1})
    =\frac
    {
    p'(\vartheta_i \mid \mathbf x, c, \boldsymbol \vartheta_{1:i-1})
    }{
    p(\vartheta_i)
    }
\end{equation}
The distribution of our auxiliary model is now given by
\begin{multline}
\vartheta_i, \mathbf x, c, \boldsymbol \vartheta_{1:i-1} \sim 
p'(\mathbf x , c, \vartheta_i, \boldsymbol \vartheta_{1:i-1}) \\
\equiv
p'(\mathbf x \mid c, \vartheta_i, \boldsymbol \vartheta_{1:i-1})p(c) 
p(\boldsymbol \vartheta_{1:i})\;,
\end{multline}
where we make use of the definitions
$p'(\mathbf x \mid c = -1, \vartheta_i, \boldsymbol \vartheta_{1:i-1}) \equiv p(\mathbf x)$  and
$p'(\mathbf x \mid c = 1, \vartheta_i, \boldsymbol \vartheta_{1:i-1}) \equiv
p(\mathbf x \mid \vartheta_i, \boldsymbol \vartheta_{1:i-1})$.

Once the above ratio estimator is trained, we can obtain the desired conditional posterior-to-conditional prior ratio by using
\begin{equation}
\frac{
p(\vartheta_i \mid \mathbf x, \boldsymbol \vartheta_{1:i-1})
}{
p(\vartheta_i \mid \boldsymbol \vartheta_{1:i-1})
}
\approx
\frac{
r(\vartheta_i ; \mathbf x, c = +1, \boldsymbol \vartheta_{1:i-1})
}{
r(\vartheta_i ; \mathbf x, c = -1, \boldsymbol \vartheta_{1:i-1})
}\;.
\end{equation}

The advantage of this formulation is that it reduces the number of networks to train by a factor of two, which should correspondingly reduce GPU memory requirements and training time. A potential downside of this approach is that the network capacity has to be high enough in order to efficiently learn both posterior and prior approximations at the same time. We will leave a quantitative comparison between different methods and network architectures to future work.

\section{Multivariate Gaussian experiment}
\label{apx:toy}
\begin{figure*}
\includegraphics[width=\linewidth]{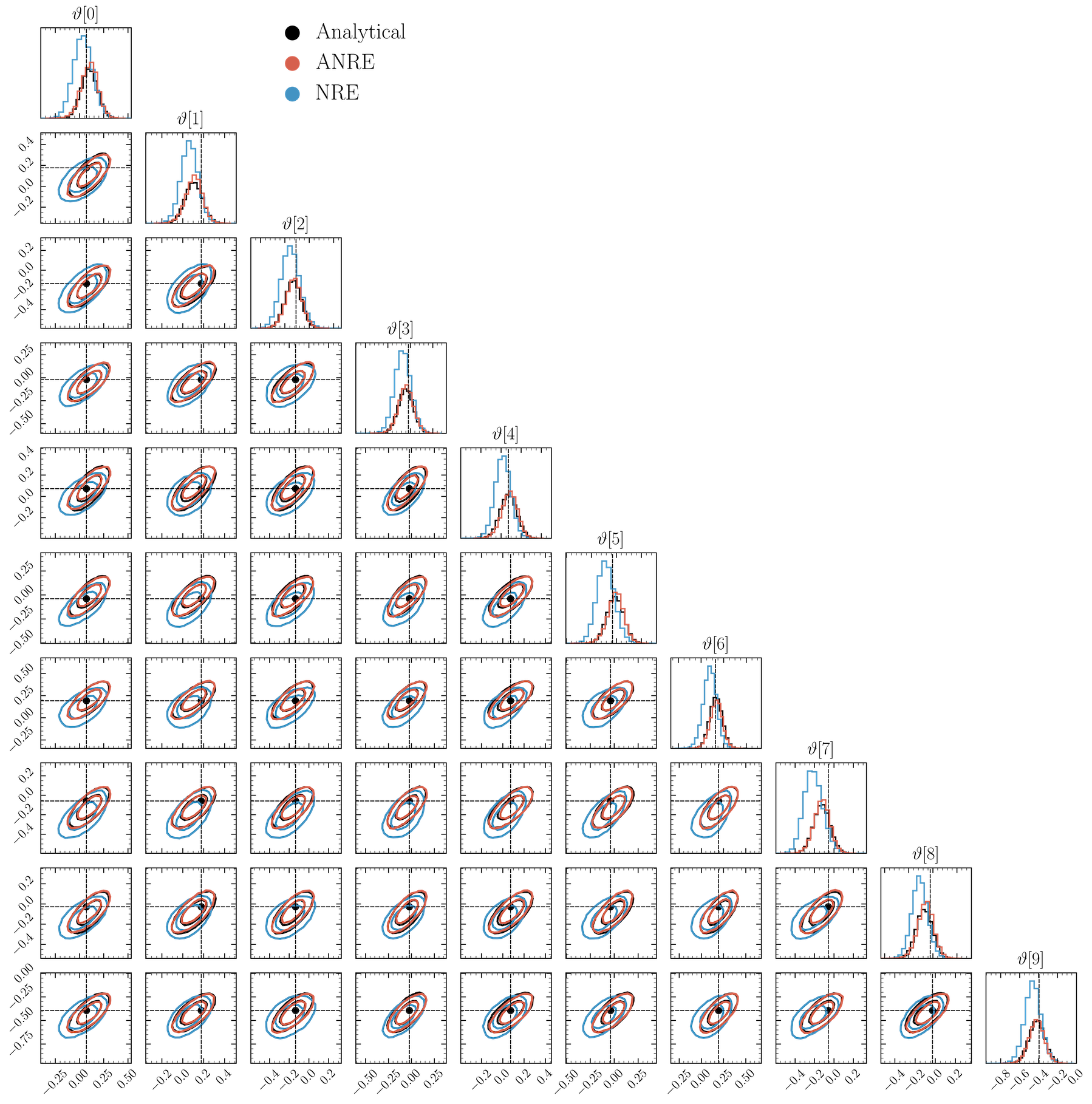}
\caption{
 Multivariate Gaussian toy example results. Corner plot highlighting the agreement between the analytic and estimated posteriors for the $d = 10$ Gaussian model described in Sec.~\ref{toy}.}
\label{fig:toy_corner}
\end{figure*}

\noindent In Fig.~\ref{fig:toy_corner}, we show the agreement between the analytical and estimated posteriors for the NRE and ANRE methods applied to the toy problem described in Sec.~\ref{toy}. We see that across all $10$ parameters the autoregressive model achieves almost perfect agreement with the analytic result.

\begin{figure*}
\includegraphics[width=\linewidth]{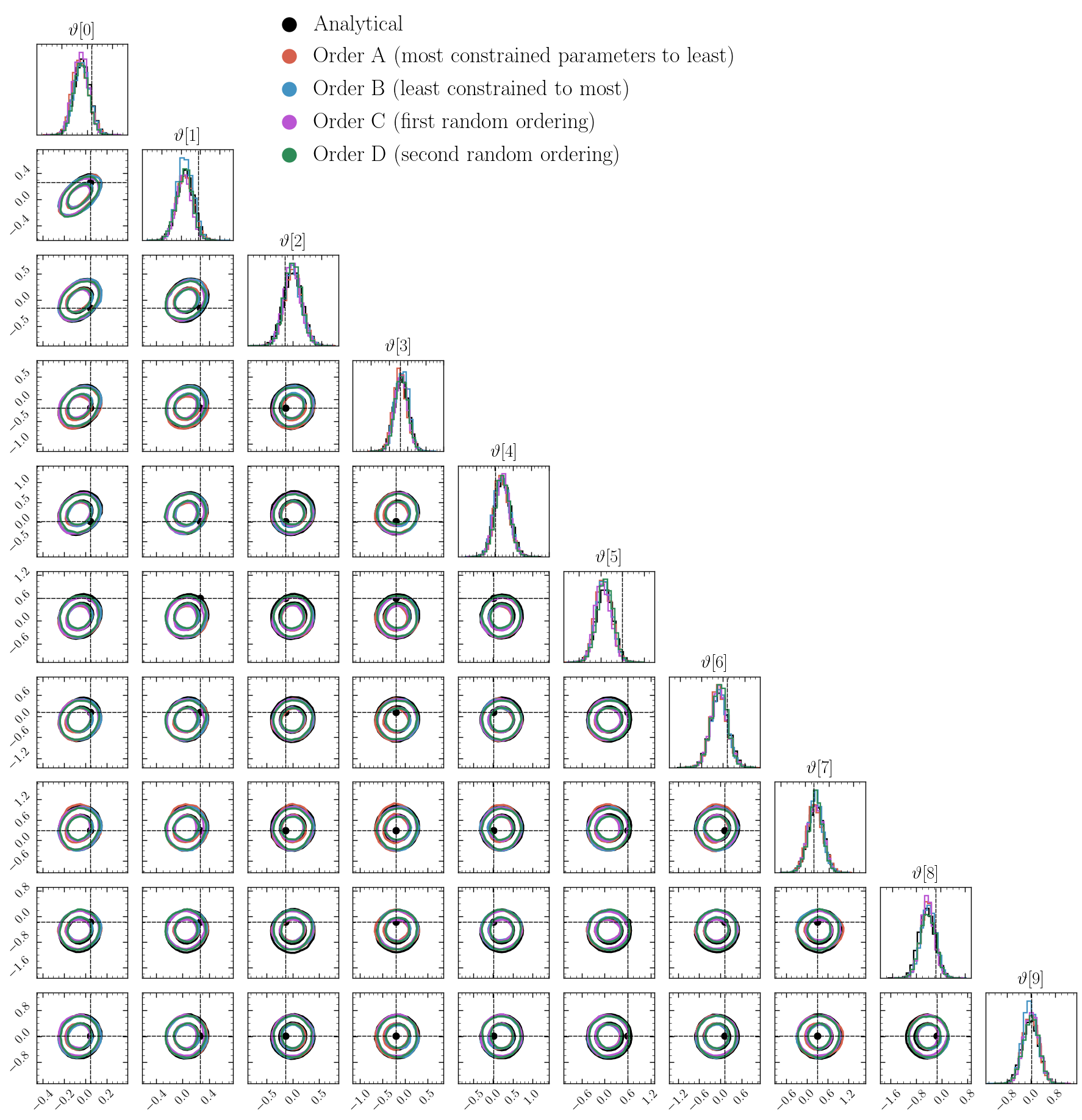}
\caption{
 Multivariate Gaussian toy example results. Corner plot highlighting the agreement between the analytic and estimated posteriors for the $d = 10$ Gaussian model described in Sec.~\ref{toy} and App.~\ref{apx:toy}. For the estimated posteriors we have used four different orderings, as described in App.~\ref{apx:toy}.}
\label{fig:toy_ordering}
\end{figure*}

In order to check the impact of the autoregressive ordering in this Multivariate Gaussian toy example we have conducted an additional test for $10$ dimensions. For this test, we use a likelihood with fixed covariance $\mathbf{\Sigma}$, which has correlation scales of $0.1$ for the off-diagonal entries, and varying diagonal correlation scales, ranging from $0.1$ to $0.55$ in $0.05$ steps. We have run our autoregressive model for four different variable orderings: from most constrained to least, from least constrained to most, and two random orderings. The results shown in Fig.~\ref{fig:toy_ordering} demonstrate that the autoregressive ordering does not have an impact in this case.

\begin{table}
    \centering
    \renewcommand{\arraystretch}{1.2}
    \begin{tabular}{l c c c c r}
        \hline
        Parameter & True value & Initial prior  \\
        \hline
        $t_\mathrm{age}\,\mathrm{[Myr]}$ & 3000 & $\mathcal{U}(2583, 3647)$ \\
        $\lambda_\mathrm{rel}$ & 1.405 & $\mathcal{U}(0.13, 2.00)$ \\
        $\lambda_\mathrm{match}$ & 1.846 & $\mathcal{U}(0.31, 2.00)$ \\
        $\xi_0$ & 0.001 & $\mathcal{U}(0.0001,0.01)$ \\
        $\alpha$ & 20.9 & $\mathcal{U}(10.0, 30.0)$ \\
        $r_h \, \mathrm{[pc]}$ & 0.001 & $\mathcal{U}(0.0001, 0.01)$ \\
        $\bar{m}\,\mathrm{[M}_\odot\mathrm{]}$ & 3 & $\mathcal{U}(1.0, 20.0)$ \\
        $\log(M_\mathrm{sat}/\mathrm{M}_\odot)$ & 4.05 & $\mathcal{U}(3.3, 4.5)$ \\
        $\sigma_v \,\mathrm{[km/s]}$ & 1.1 & $\mathcal{U}(0.96, 1.32)$ \\
        $x_c\,\mathrm{[kpc]}$ & 11.8 & $\mathcal{U}(117, 119)$ \\
        $y_c\,\mathrm{[kpc]}$ & 0.79 & $\mathcal{U}(0.6, 1.1)$ \\
        $z_c\,\mathrm{[kpc]}$ & 6.4 & $\mathcal{U}(6.32, 6.54)$ \\
        $v_{x,c}\,\mathrm{[km/s]}$ & 109.5 & $\mathcal{U}(106.8, 113.7)$ \\
        $v_{y,c}\,\mathrm{[km/s]}$ & -254.5 & $\mathcal{U}(-256, -251)$ \\
        $v_{z,c}\,\mathrm{[km/s]}$ & -90.3 & $\mathcal{U}(-93.3,-84.6)$ \\
        $p_\mathrm{near}$ & 0.5 & $\mathcal{U}(0.34, 0.72)$ \\
        \hline
    \end{tabular}
    \caption{True parameter values and priors used in the TMNRE inference round for the stellar streams example. Note that our prior choices are taken from the final round of inference of \protect\cite{alvey2023albatross} so as to test the autoregressive model in a final round of precision inference through active learning.}
    \label{tab:streams-params}
\end{table}

\section{Stellar streams experiment}
\label{apx:streams}

\noindent In this short appendix we present the technical details relevant for the stellar streams experiment presented in Sec.~\ref{subsec:stream}. There are a number of things directly analogous to the implementation described in \cite{alvey2023albatross}. In particular, we use the stellar streams simulator, \texttt{sstrax} \citep{2023ascl.soft06008A}, developed in this reference, as well as the \texttt{swyft}-based inference code \texttt{albatross} \citep{2023ascl.soft06009A}. In addition, we consider the same set of model parameters (described in Tab.~\ref{tab:streams-params} and in detail in \cite{alvey2023albatross}), the same injection parameters to generate the observation, and the same binning/observation scheme. The key difference with respect to this reference is in the inference network used. In particular, we replace the $1$-dimensional ratio estimation across all model parameters with the autoregressive method described in this work to model their joint distribution. In terms of network training details, we use the Adam optimiser with an initial learning rate of $10^{-6}$. In addition, we use a batch size of $512$ and a simulation budget of $3\times 10^5$ training examples. Finally, we use the following orderings (Order A and Order B referenced in the main text) for training the autoregressive ratio estimator: [$\sigma_v$, $x_c$, $y_c$, $z_c$, $v_{x,c}$, $v_{y, c}$, $v_{z, c}$, $t_\mathrm{age}$, $p_\mathrm{near}$, $\lambda_\mathrm{rel}$, $\lambda_{match}$, $\xi_0$, $\alpha$, $r_h$, $\bar{m}$, $\log(M_\mathrm{sat}/\mathrm{M}_\odot)$] (Order A), and [$\xi_0$, $\lambda_{match}$, $\lambda_\mathrm{rel}$, $\alpha$, $r_h$, $\bar{m}$, $\log(M_\mathrm{sat}/\mathrm{M}_\odot)$, $t_\mathrm{age}$, $\sigma_v$, $x_c$, $y_c$, $z_c$, $v_{x,c}$, $v_{y, c}$, $v_{z, c}$, $p_\mathrm{near}$] (Order B).

\begin{figure*}
    \includegraphics[width=\linewidth]{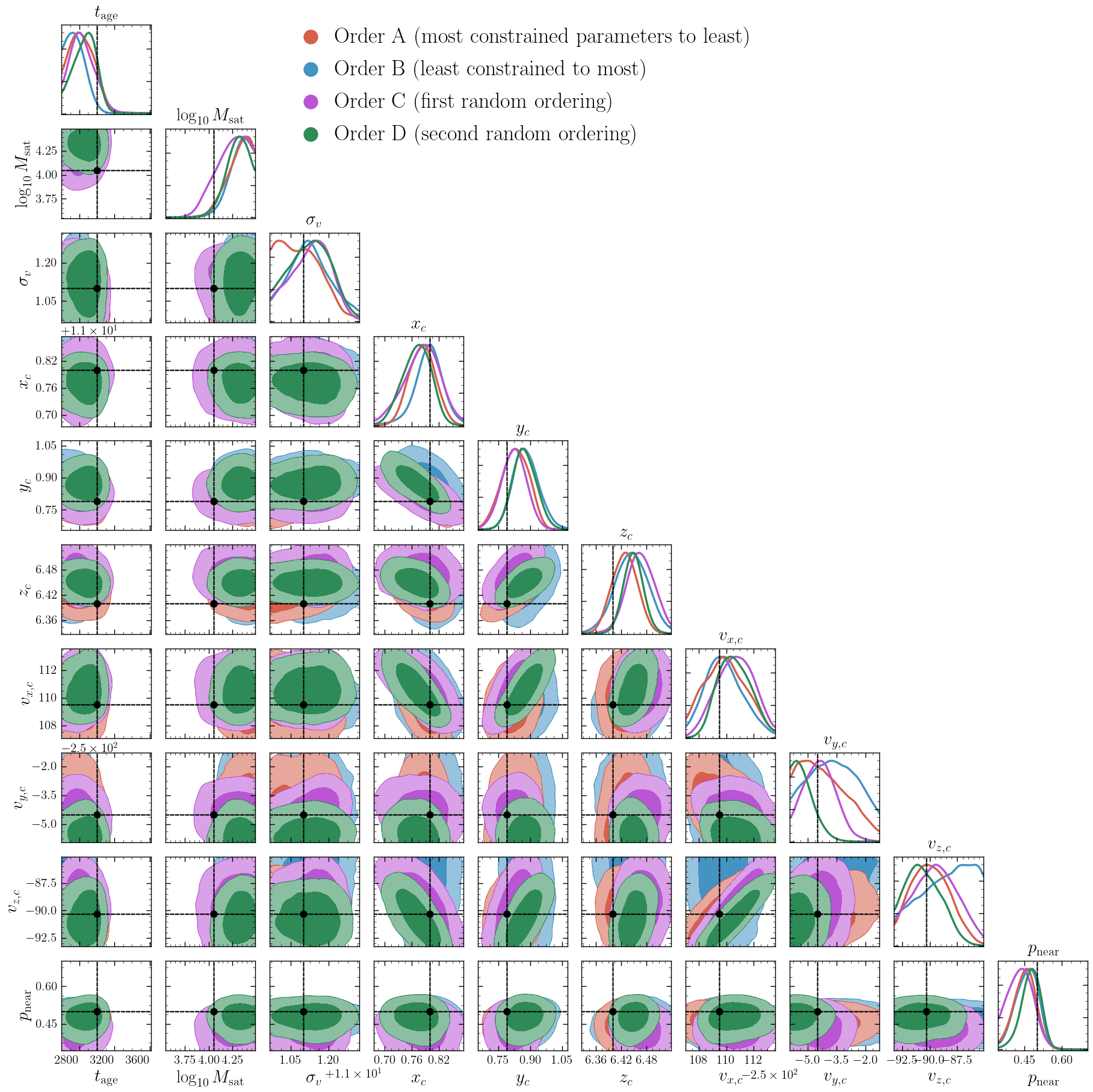}
    \caption{Additional checks of the variable ordering conclusions given in Sec.~\ref{subsec:stream}. Here, we overlay the results presented in the main text with inference results given two other random orderings.}
    \label{fig:corner_streams_2}
\end{figure*}

In Fig.~\ref{fig:corner_streams_2}, we also show the results of a test with two additional random orderings. Specifically, the orderings shown are [$\alpha$, $\log(M_\mathrm{sat}/\mathrm{M}_\odot)$, $\lambda_\mathrm{rel}$, $\xi_0$, $v_{x,c}$, $\sigma_v$, $v_{z,c}$, $\bar{m}$, $t_\mathrm{age}$, $r_h$, $x_c$, $z_c$, $p_\mathrm{near}$, $\lambda_\mathrm{match}$, $y_c$, $v_{y,c}$] (first random order) and [$v_{z,c}$, $p_\mathrm{near}$, $\bar{m}$, $\sigma_v$, $v_{y,c}$, $\alpha$, $v_{x, c}$, $t_\mathrm{age}$, $\lambda_{\mathrm{rel}}$, $z_c$, $\lambda_\mathrm{match}$, $y_c$, $x_c$, $\xi_0$, $r_h$, $\log(M_\mathrm{sat}/\mathrm{M}_\odot)$] (second random order).

\begin{figure}
    \centering
    \includegraphics[width=\linewidth]{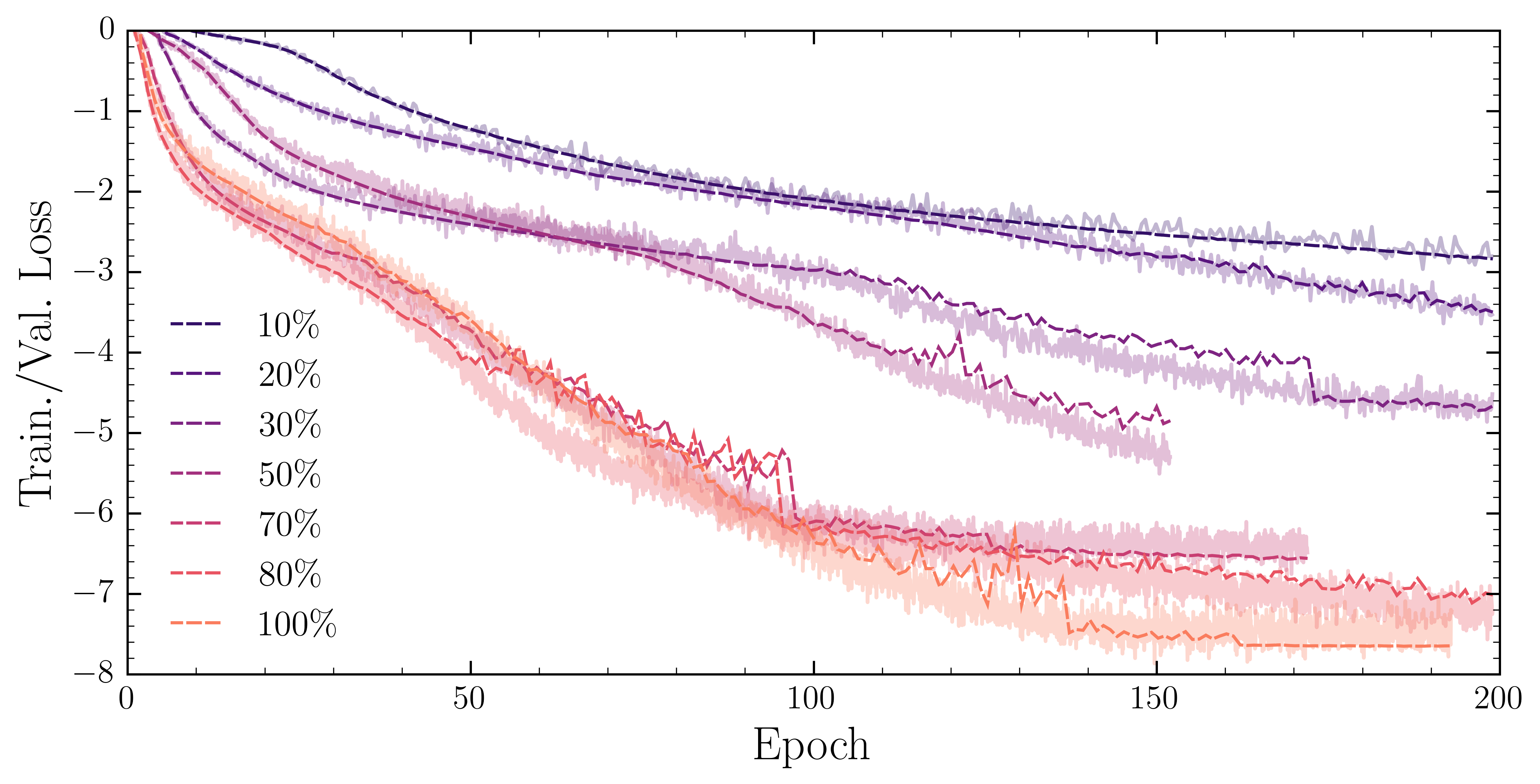}
    \caption{Training loss for order B (see Fig.~\ref{fig:corner_streams_2} and Sec.~\ref{subsec:stream} and Fig.~\ref{fig:streams_loss} in the main text) as a function of epochs for different simulation budgets, with a maximum budget of $6\times 10^5$ simulations. The validation loss for each case is shown as a dashed line in the figure.}
    \label{fig:anre-training}
\end{figure}

Furthermore, we performed an additional test in order to check when the loss for order B reaches the one for order A in Fig.~\ref{fig:streams_loss} (see main text for definitions), by gradually increasing the simulation budget.
Fig.~\ref{fig:anre-training} shows the loss for order B as a function of epochs, for different sizes of training dataset, with the biggest one (100\% in the figure) containing $6\times10^5$ simulations. From Fig.~\ref{fig:anre-training}, we can see that the loss for order B reaches a plateau value of -6.1 (the one of the loss for order A in Fig.~\ref{fig:streams_loss} when training on $3\times10^5$ simulations) for approximately between 60\% to 70\% of the total $6\times10^5$ simulations, so roughly for $4\times10^5$ simulations. In this specific example, this experimental check shows that a network trained with order B needs 25\% more simulations to plateau at the same loss as one trained for order A. This test further confirms our observation that the ‘penalty’ for choosing a non-optimal ordering is not severe, and can be easily circumvented, for example as in this case, by increasing the simulation budget. However, the important take away message is that the simulation computational budget can be reduced by choosing an optimal ordering. 

\section{Gravitational lensing experiment}
\label{apx:lensing}

\noindent In this appendix we provide a more detailed account of the model adopted to generate the strong lensing simulations, the training details, and the employed neural networks to obtain the results presented in Sec.~\ref{subsec:lensing}.

\subsection{Simulator}

\begin{table}
    \centering
    \renewcommand{\arraystretch}{1.2}
    \begin{tabular}{c c c c c c c}
        \hline
        Component & Parameter & True value & Initial prior  \\
        \hline
        \parbox[t]{1mm}{\multirow{3}{*}{\rotatebox[origin=c]{90}{Subhalo}}}
        & $x_\mathrm{sub}\, ['']$ & -1.2 & $\mathcal{U}(-2.5, 2.5)$ \\
        & $y_\mathrm{sub}\, ['']$ & 1 & $\mathcal{U}(-2.5, 2.5)$ \\
        & $\log_{10} m_\mathrm{sub}\, [M_\odot]$ & $9.5$ & $\mathcal{U}(8, 11)$ \\
        \hline
        \parbox[t]{1mm}{\multirow{6}{*}{\rotatebox[origin=c]{90}{SPLE}}}
        & $x_\mathrm{lens}\, ['']$ & 0.1 & $\mathcal{U}(-0.2, 0.2)$  \\
        & $y_\mathrm{lens}\, ['']$ & 0.05 & $\mathcal{U}(-0.2, 0.2)$ \\
        & $\varphi_\mathrm{lens} \, [^\circ]$ & 0.3 & $\mathcal{U}(0, 1.5)$ \\
        & $q_\mathrm{lens}$ & 0.89 & $\mathcal{U}(0.1, 1)$ \\
        & $\gamma$ & 2 & $\mathcal{U}(1.8, 2.2)$ \\
        & $r_\mathrm{ein}\, ['']$ & 1.5 & $\mathcal{U}(1, 2)$ \\
        \hline
        \parbox[t]{1mm}{\multirow{2}{*}{\rotatebox[origin=c]{90}{Shear}}}
        & $\gamma_1$ & 0.01 & $\mathcal{U}(-0.05, 0.05)$  \\
        & $\gamma_2$ & -0.02 & $\mathcal{U}(-0.05, 0.05)$ \\
        \hline
        \parbox[t]{1mm}{\multirow{7}{*}{\rotatebox[origin=c]{90}{Lens light}}}
        & $x_\mathrm{light}\, ['']$ & 0.03 & $\mathcal{U}(-0.1, 0.1)$ \\
        & $y_\mathrm{light}\, ['']$ & 0.02 & $\mathcal{U}(-0.1, 0.1)$ \\
        & $\varphi_\mathrm{light} \, [^\circ]$ & 0.3 & $\mathcal{U}(0., 1.5)$ \\
        & $q_\mathrm{light}$ & 0.7 & $\mathcal{U}(0.1, 1)$ \\
        & $n$ & 1.58 & $\mathcal{U}(0.1, 4)$ \\
        & $r_e\, ['']$ & 2.1 & $\mathcal{U}(0.1, 3)$ \\
        & $I_e$ & 1.2 & $\mathcal{U}(0, 4)$ \\
        \hline
        \parbox[t]{1mm}{\multirow{7}{*}{\rotatebox[origin=c]{90}{Source light}}}
        & $x_\mathrm{src}\, ['']$ & 0.02 & $\mathcal{U}(-0.2, 0.2)$ \\
        & $y_\mathrm{src}\, ['']$ & 0.08 & $\mathcal{U}(-0.2, 0.2)$ \\
        & $\varphi_\mathrm{src} \, [^\circ]$ & 0.7 & $\mathcal{U}(0., 1.5)$ \\
        & $q_\mathrm{src}$ & 0.8 & $\mathcal{U}(0.1, 1)$ \\
        & $n$ & 1.5 & $\mathcal{U}(0.1, 4)$ \\
        & $r_e\, ['']$ & 1.4 & $\mathcal{U}(0.1, 3)$ \\
        & $I_e$ & 2.5 & $\mathcal{U}(0, 4)$ \\
        \hline
    \end{tabular}
    \caption{True subhalo and macro-model parameter values and priors used in the first TMNRE inference round in the strong gravitational lensing application. This prior generates images with variance displayed in the first row of Fig.~\ref{fig:targeted_data}.}
    \label{tab:lensing-params}
\end{table}

\noindent We use the simulator adopted in \cite{coogan2020targeted, coogan2022walks, Anau_Montel_2022} to generate strong lensing image observations. To model the lens light and the source light flux we use a Sèrsic profile \citep{Sersic_1963}, parameterized by seven variables: position, position angle, axis ratio, index, effective radius, and surface intensity. We adopt a singular power-law ellipsoid (SPLE) \citep{Suyu_2009} for the main lens mass distribution, with a total of six parameters: position on the lens plane, position angle, axis ratio, slope, and Einstein radius. We consider two additional parameters to model the external shear. In total, there are twenty-two macro-model parameters. To model the density profile of the dark matter subhalo we adopt the smoothly truncated universal Navarro-Frenk-White mass density profile from \cite{Baltz_2009}. We fix the truncation radius to $\tau=6$. The subhalo is then described by three parameters: virial mass $m_\mathrm{sub}$, and position on the lens plane $(x_\mathrm{sub}, y_\mathrm{sub})$. We adopt the concentration-mass relation from \cite{Correa_2015}. We show each parameter prior and the value with which we have generated our mock target observation in Tab.~\ref{tab:lensing-params}.

Given this model, we generate $100 \times 100$ pixel$^2$ images with a resolution of $0.05"$ per pixel side, for a total field of view of $5" \times 5"$ in an image. The instrumental effects include a Gaussian point spread function with a full width at half maximum of $0.05"$ and Gaussian noise. We choose redshift $z_\mathrm{lens}=0.9$ for the lens and $z_\mathrm{source}=2$ for the source.

\subsection{Training details and neural networks}

\begin{table}
    \centering
    \begin{tabular}{c}
        \hline
        \texttt{Conv2d(1, 4, 3, 2, 1, bias=True)} \\
        \texttt{BatchNorm2d(4)} \\
        \texttt{ReLU} \\
        \hline
        \texttt{Conv2d(4, 8, 5, 2, 1, bias=True)} \\
        \texttt{BatchNorm2d(8)} \\
        \texttt{ReLU} \\
        \hline
        \texttt{Conv2d(8, 16, 5, 2, 1, bias=True)} \\
        \texttt{BatchNorm2d(16)} \\
        \texttt{ReLU} \\
        \hline
        \texttt{Conv2d(16, 32, 5, 2, 1, bias=True)} \\
        \texttt{BatchNorm2d(32)} \\
        \texttt{ReLU} \\
        \hline
        \texttt{Flatten()} \\
        \texttt{LazyLinear(256)} \\
        \hline
    \end{tabular}
    \caption{The convolutional compression network used in the macro-model parameter ratio estimator. The notation is taken from \texttt{PyTorch}: the arguments to \texttt{Conv2d} are the number of input channels, output channels, kernel size, stride and padding, respectively. The horizontal lines highlight where the number of channels changes. Note that we standardize the images before providing them to the convolutional networks.}
    \label{tab:macro}
\end{table}
\begin{table}
    \centering
    \begin{tabular}{r l}
        \hline
        \texttt{image\_size} & \texttt{100} \\
        \texttt{n\_channels} & \texttt{1} \\
        \texttt{n\_classes} & \texttt{8}\\
        \texttt{s} & \texttt{1} \\
        \hline
    \end{tabular}
    \caption{The details of the UNet network used in the subhalo. We use the implementation from \url{https://github.com/milesial/Pytorch-UNet}, with arguments given in the table. The UNet output is then sampled at the positions of the subhalo contrastive examples to bring it to a lower dimensionality, and this feature vector is passed to the binary classifier.
    }
    \label{tab:UNet}
\end{table}

\noindent For all tasks we use the same general ratio estimator architecture. It consists of an initial compression network that maps the high-dimensional lensing observation $\mathbf x$ into a feature vector. This feature vector is concatenated to the parameters we want to infer. The vector is then passed to  \texttt{swyft} binary classifier which outputs an estimate of the likelihood-to-evidence ratio. The compressor architecture for the macro-model parameters is given in Tab.~\ref{tab:macro}. This same compression is used when estimating 1-dimensional marginal ratio estimators, or the joint one with NRE or autoregressive NRE, as explained in Sec.~\ref{subsec:lensing}. The compressor architecture for the subhalo parameters is given in Tab.~\ref{tab:UNet}.

We used the Adam optimizer with an initial learning rate of $10^{-3}$ for the macro-model ratio estimator and $8 \times 10^{-5}$ for the subhalo ratio estimator, and a batch size of $64$. The learning rate was reduced by a factor of $0.1$ whenever the validation loss plateaued for $2$ epochs. Training was run until the validation loss stopped improving for more than $5$ epochs. The results shown in Fig.~\ref{fig:corner} are obtained using $2\times 10^5$ training samples. To run the inference, we use the model weights obtained at the lowest validation loss curve point.


\label{lastpage}
\end{document}